\documentclass[twocolumn,trackchanges]{aastex63}

\usepackage{color, xcolor, xspace, todonotes, comment}
\usepackage[normalem]{ulem}
\newcommand\redsout{\bgroup\markoverwith{\textcolor{red}{\rule[0.5ex]{2pt}{0.4pt}}}\ULon}

\newcommand{\gal}{\mbox{ESO 253$-$G003}\xspace}
\newcommand{\asassn}{\mbox{ASASSN-14ko}\xspace}

\renewcommand{\vr}{\hbox{$v_r$}\xspace}
\newcommand{\vfwhm}{\hbox{$v_{\rm{FWHM}}$}\xspace}
\newcommand{\OIII}{\hbox{[\ion{O}{3}]}\xspace}

\newcommand{\OIIIb}{\hbox{[\ion{O}{3}]~$\lambda5007$}\xspace}
\newcommand{\OIIIlong}{\hbox{[\ion{O}{3}]~$\lambda\lambda4959,5007$}\xspace}

\newcommand{\SII}{\hbox{[\ion{S}{2}]}\xspace}

\newcommand{\SIIlong}{\hbox{[\ion{S}{2}]~$\lambda\lambda6716,6731$}\xspace}
\newcommand{\Ha}{\hbox{H$\alpha$}\xspace}
\newcommand{\Hb}{\hbox{H$\beta$}\xspace}
\newcommand{\OI}{\hbox{[\ion{O}{1}]}\xspace}
\newcommand{\OIlong}{\hbox{[\ion{O}{1}]~$\lambda\lambda6300,6363$}\xspace}

\newcommand{\NII}{\hbox{[\ion{N}{2}]}\xspace}
\newcommand{\NIIa}{\hbox{[\ion{N}{2}]~$\lambda6548$}\xspace}
\newcommand{\NIIb}{\hbox{[\ion{N}{2}]~$\lambda6583$}\xspace}
\newcommand{\NIIlong}{\hbox{[\ion{N}{2}]~$\lambda\lambda6548,6583$}\xspace}

\newcommand{\HII}{\hbox{\ion{H}{2}}\xspace}

\newcommand{\HeIIlong}{\hbox{\ion{He}{2}~$\lambda4686$}\xspace}
\newcommand{\otwo}{\hbox{O$_2$}\xspace}

\newcommand{\ArIII}{\hbox{[\ion{Ar}{3}]}\xspace}
\newcommand{\CaII}{\hbox{\ion{Ca}{2}}\xspace}
\newcommand{\FeVII}{\hbox{[\ion{Fe}{7}]}\xspace}
\newcommand{\FeVIIa}{\hbox{[\ion{Fe}{7}]~$\lambda5720$}\xspace}
\newcommand{\FeVIIb}{\hbox{[\ion{Fe}{7}]~$\lambda6086$}\xspace}
\newcommand{\logSIIHa}{\hbox{$\log_{10}$(\SII/\Ha)}\xspace}
\newcommand{\logOIHa}{\hbox{$\log_{10}$(\OI/\Ha)}\xspace}
\newcommand{\logOIIIHb}{\hbox{$\log_{10}$(\OIII/\Hb)}\xspace}
\newcommand{\logNIIHa}{\hbox{$\log_{10}$(\NII/\Ha)}\xspace}

\newcommand{\kms}{\hbox{km~s$^{-1}$}\xspace}
\newcommand{\gaia}{$Gaia$\xspace}

\submitjournal{MNRAS}

\shorttitle{MUSE Data of ASASSN-14ko}
\shortauthors{Tucker et al.}
\graphicspath{{./}{figures/}}
\begin{document}

\title{An AMUSING Look at the Host of the Periodic Nuclear Transient ASASSN-14ko \\ Reveals a Second AGN}

%\title{MUSE Data of Periodic Galaxy Transient ASASSN-14ko Reveal a Second AGN}

\correspondingauthor{Michael Tucker}
\email{tuckerma95@gmail.com}

\author[0000-0002-2471-8442]{M. A. Tucker}
\altaffiliation{DOE CSGF Fellow}
\affiliation{Institute for Astronomy, University of Hawaii at Manoa, 2680 Woodlawn Dr., Honolulu, HI 96822}

\author[0000-0003-4631-1149]{B. J. Shappee}
\affiliation{Institute for Astronomy, University of Hawaii at Manoa, 2680 Woodlawn Dr., Honolulu, HI 96822}

\author[0000-0001-9668-2920]{J. T. Hinkle}
\affiliation{Institute for Astronomy, University of Hawaii at Manoa, 2680 Woodlawn Dr., Honolulu, HI 96822}

\author[0000-0001-7351-2531]{J. M. M. Neustadt}
\affiliation{Department of Astronomy, The Ohio State University, 140 West 18th Avenue, Columbus, OH 43210, USA}

\author[0000-0002-3719-940X]{M. Eracleous}
\affiliation{Department of Astronomy \& Astrophysics and Institute for Gravitation and the Cosmos, The Pennsylvania State University, 525 Davey Laboratory, University Park, PA 16802}

\author[0000-0001-6017-2961]{C.~S.~Kochanek}
\affiliation{Department of Astronomy, The Ohio State University, 140 West 18th Avenue, Columbus, OH 43210, USA}
\affiliation{Center for Cosmology and AstroParticle Physics, The Ohio State University, 191 W. Woodruff Ave., Columbus, OH 43210, USA}

\author[0000-0003-1072-2712]{J.~L.~Prieto}
\affiliation{N\'ucleo de Astronom\'ia de la Facultad de Ingenier\'ia y Ciencias, Universidad Diego Portales, Av. Ej\'ercito 441, Santiago, Chile}
\affiliation{Millennium Institute of Astrophysics, Santiago, Chile}

\author[0000-0003-3490-3243]{A. V. Payne}
\altaffiliation{NASA Fellow}
\affiliation{Institute for Astronomy, University of Hawaii at Manoa, 2680 Woodlawn Dr., Honolulu, HI 96822}

\author[0000-0002-1296-6887]{L. Galbany}
\affiliation{Institute of Space Sciences (ICE, CSIC), Campus UAB, Carrer de Can Magrans, s/n, E-08193 Barcelona, Spain}

\author[0000-0003-0227-3451]{J. P. Anderson}
\affiliation{European Southern Observatory, Alonso de C\'ordova 3107, Casilla 19, Santiago, Chile}

\author[0000-0002-4449-9152]{K.~Auchettl}
\affiliation{School of Physics, The University of Melbourne, Parkville, VIC 3010, Australia}
\affiliation{ARC Centre of Excellence for All Sky Astrophysics in 3 Dimensions (ASTRO 3D)}
\affiliation{Department of Astronomy and Astrophysics, University of California, Santa Cruz, CA 95064, USA}

\author[0000-0002-5504-8752]{C. Auge}
\affiliation{Institute for Astronomy, University of Hawaii at Manoa, 2680 Woodlawn Dr., Honolulu, HI 96822}

\author[0000-0001-9206-3460]{Thomas~W.-S.~Holoien}
\altaffiliation{NHFP Einstein Fellow}
\affiliation{The Observatories of the Carnegie Institution for Science, 813 Santa Barbara St., Pasadena, CA 91101, USA}


\begin{abstract}

We present Multi-Unit Spectroscopic Explorer (MUSE) integral-field spectroscopy of \gal, which hosts a known Active Galactic Nucleus (AGN) and the periodic nuclear transient \asassn, observed as part of the All-weather MUse Supernova Integral-field of Nearby Galaxies (AMUSING) survey. The MUSE observations reveal that the inner region hosts two AGN separated by $1.4\pm0.1~\rm{kpc}$ ($\approx 1\farcs7$). The brighter nucleus has asymmetric broad, permitted emission-line profiles and is associated with the archival AGN designation. The fainter nucleus does not have a broad emission-line component but exhibits other AGN characteristics, including $\vfwhm\approx 700~\kms$ forbidden line emission, $\logOIIIHb \approx 1.1$, and high excitation potential emission lines such as \FeVIIb and \HeIIlong. The host galaxy exhibits a disturbed morphology with large kpc-scale tidal features, potential outflows from both nuclei, and a likely superbubble. A circular relativistic disk model cannot reproduce the asymmetric broad emission-line profiles in the brighter nucleus, but two non-axisymmetric disk models provide good fits to the broad emission-line profiles: an elliptical disk model and a circular disk + spiral arm model. Implications for the periodic nuclear transient \asassn are discussed.
\\

\end{abstract}



\section{Introduction} \label{sec:intro}

Galaxy mergers are a natural consequence of $\Lambda$ Cold Dark Matter ($\Lambda$CDM) cosmology (see \citealp{somerville2015} and references therein). Most or all galaxies host supermassive black holes (SMBHs) at their centers \citep[e.g., ][]{kormendy2013} leading to SMBH interactions and mergers through cosmic time \citep[e.g., ][]{begelman1980}. Merging galaxies present an ideal location for finding multiple SMBH systems \citep[e.g., ][]{menou2001, cuadra2009, khan2012} and for understanding both their evolution and their effect on the surrounding stellar environment \citep[e.g., ][]{kormendy2013}. Galactic mergers stir gas and dust \citep[e.g., ][]{springel2000}, increasing the likelihood of ``feeding'' SMBHs with gas and increasing the prevalence of Active Galactic Nuclei (AGN) in these systems \citep[e.g., ][]{hopkins2006, treister2012, goulding2018}. 

AGN are thought to play an important role in galaxy evolution \citep[see][for a review]{kormendy2013}. The hard radiation spectrum photoionizes the surrounding gas, producing distinctive emission-line ratios \citep[e.g., ][]{veilleux1987} and regulating accretion onto the SMBH \citep[e.g., ][]{ciotti2007, park2012}. The accretion also drives outflows and/or jets into the host-galaxy \citep[e.g., ][]{silk1998, ciotti2001, harrison2014, nardini2018} which interact with the galactic interstellar medium (ISM) and regulate star-formation (SF) \citep[e.g., ][]{hopkins2006, dubois2012, zubovas2013}. 

Merging galaxies provide a unique avenue for understanding AGN formation, especially the formation of multiple AGN systems \citep[e.g., ][]{fu2011, comerford2015, gabanyi2016, liu2019, benitez2019, kollatschny2020}. Systematic searches find higher AGN occurrence rates in merging systems than in the general field population \citep[e.g., ][]{surace1998, schmitt2001, ellison2011, gao2020, secrest2020}, although the results can depend on the AGN selection process \citep[e.g., ][]{ellison2015}.

Interestingly, merging galaxies may also provide a new location for enhanced production rates of tidal disruption events (TDEs). Binary SMBHs are thought to increase the galactic TDE rate by up to an order of magnitude, usually through three-body interactions \citep[e.g., ][]{ivanov2005, chen2009, wegg2011, liu2013, li2019}. Several TDE candidates have been discovered in merging systems \citep[e.g., ][]{tadhunter2017, mattila2018, kool2020}, although dust obscuration complicates occurrence rate analyses. A similar enhancement of TDE production has been observed in post-starburst or ``E+A'' galaxies \citep[e.g., ][]{arcavi2014, prieto2016, french2016, french2020} which also exhibit signs of galactic mergers \citep[e.g., ][]{zabludoff1996, pawlik2018, chen2019}. The increase in AGN and TDE rates is consistent with galactic mergers augmenting nuclear densities and modifying stellar dynamics in these systems, refilling the SMBH loss cone(s) \citep[e.g., ][]{merritt2005, khan2011, kelley2017} and boosting the chances of stellar and gaseous interactions with the SMBH(s) \citep[e.g., ][]{mihos1996, vanwassenhove2014, capelo2017}.

Here we explore \gal with Multi-Unit Spectroscopic Explorer (MUSE; \citealp{bacon2010}) integral-field unit (IFU) spectroscopy. \gal exhibits many features of a late-stage merger including two nuclei and kpc-scale tidal arms. The brighter nucleus hosts a known AGN \citep{veron-cetty2010} at $z = 0.04249\pm0.00008$ \citep{aguero1996} and the fainter, secondary nucleus has been observed in infrared (IR) imaging \citep{videla2013} $\approx 1\farcs7$ ($\sim 1.4~\rm{kpc}$) away from the brighter nucleus \citep{asmus2014}.

Recently, \citet{payne2020} reported that \asassn, a nuclear transient in \gal discovered by the All-Sky Automated Survey for SuperNovae (ASAS-SN; \citealp{shappee2014, kochanek2017}), exhibits periodic outbursts with a negative period derivative. \citet{payne2020} considered three scenarios to explain the recurrent flares and negative period derivative: a SMBH binary, a bound star interacting with the SMBH disk, and a partial TDE, which we summarize briefly here. A SMBH binary was disfavored because the predicted period derivative from gravitational wave emission is an order of magnitude smaller than the observed period derivative. The bound star scenario struggles to reproduce the stability of the flare amplitudes, shapes, and durations without fine-tuning the stellar orbit and inclination of the accretion disk to our line-of-sight. A repeating partial TDE can, for a period of time, produce periodic, self-similar flares with a period derivative consistent with the observed value without introducing new discrepancies. Thus, \citet{payne2020} favored the repeating partial TDE interpretation.

Using the MUSE data, we find that (1) the broad emission lines in the brighter nucleus cannot be explained with a circular disk model, and (2) the fainter nucleus also hosts an AGN. After outlining our data acquisition, reduction, and analysis methods in \S\ref{sec:methods}, this paper takes a ``small to large" approach. We first analyze spectra of the nuclei in \S\ref{sec:nuclei}, placing the nuclei on emission-line diagnostic diagrams and modeling the asymmetric broad emission-line profiles. Next, we analyze various galaxy-wide properties in \S\ref{sec:galaxy}, including the gas kinematics and emission-line ratios. In \S\ref{sec:discuss} we discuss models of the broad-line emission in the brighter nucleus to constrain the properties of \asassn. Finally, in \S\ref{sec:conclusion}, we summarize our findings. Throughout our analysis we use the standard cosmological parameters of $H_0 = 70 ~\rm{km}~\rm{s}^{-1}~\rm{Mpc}^{-1}$ and $\Omega_m = 0.3$ resulting in a luminosity distance of $\approx 188~\rm{Mpc}$ and a projected scale of $\approx 0.85 ~\rm{kpc}/\arcsec$.

\begin{figure*}
    \centering
    \includegraphics[width=\linewidth]{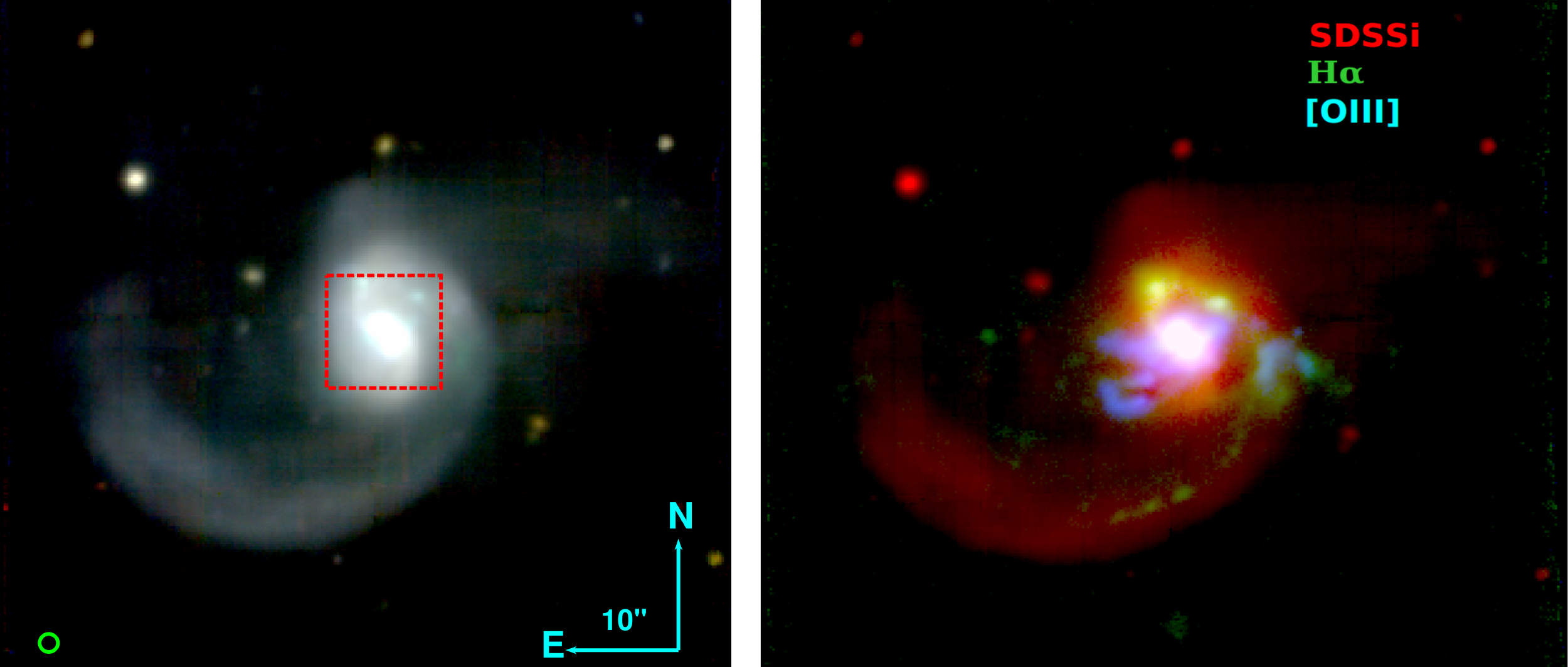}
    \caption{Broad-band (left; blue=$4800-5500~\rm\AA$, green=$6600-7300~\rm\AA$, red=$8400-9100~\rm\AA$) and emission-line (right, blue=\OIIIb, green=\Ha, red=SDSS $i$-band) images constructed from the MUSE datacube. The $10\arcsec$ image scale corresponds to a projected scale of $\approx 8.5~\rm{kpc}$ at the distance of the galaxy. The green circle in the lower left corner represents the $0\farcs8$ image FWHM measured using the bright star to the north-east of the galaxy. The dashed red square in the left panel marks the inset shown in Fig. \ref{fig:nuc_contours}.}
    \label{fig:color_images}
\end{figure*}

\begin{figure}
    \centering
    \includegraphics[width=\linewidth]{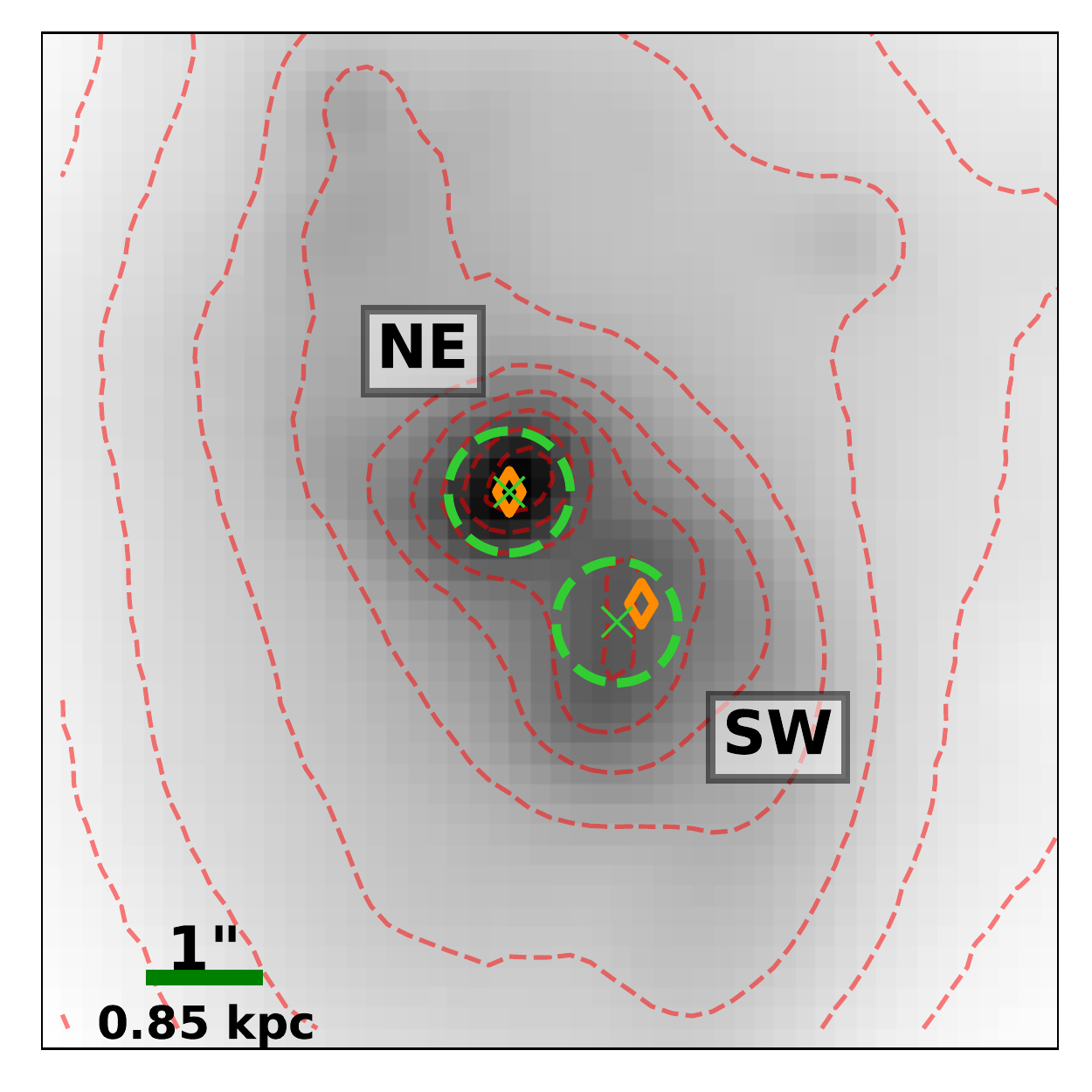}
    \caption{Zoom-in of the nuclear region of \gal (the red box in the left panel of Fig. \ref{fig:color_images}) constructed using a spectral region free from strong emission lines ($\lambda = 5100-6100~\rm\AA$). The dotted red lines are 10 logarithmically-spaced surface brightness contours. Green X's mark the centers of each nucleus computed from fitting two-dimensional Gaussian profiles and the green dashed circles represent the spectral extraction apertures used in \S\ref{sec:nuclei}. The orange diamonds mark the locations of the 
    MIR point sources \citep{asmus2014} with a small coordinate shift applied (see \S\ref{sec:nuclei}). 
    }
    \label{fig:nuc_contours}
\end{figure}

\section{Data and Methods}\label{sec:methods}

\subsection{MUSE Observations}\label{subsec:methods.obs}

Our MUSE observations of \gal were obtained on 2015-12-02 as part of the All-weather MUse Supernova Integral-field of Nearby Galaxies (AMUSING; \citealt{galbany2016, lopez-coba2020}) survey. \citet{galbany2016} describe the observing strategy and data reduction. The MUSE observations were conducted $\approx 45$~days after the previous \asassn outburst (phase $\sim 0.4$ for a period of 114~days) when the nucleus was at the quiescent optical flux level \citep{payne2020}.

The MUSE field-of-view (FoV) covers $1\arcmin\times1\arcmin$ ($\approx 51~\rm{kpc}\times51~\rm{kpc}$ at the distance of \gal) with a spatial sampling of $0\farcs2\times0\farcs2$ ($\approx0.17~\rm{kpc}\times0.17~\rm{kpc}$) per spaxel. Using the bright star to the north-east of \gal, we measure a spatial Full-Width at Half Maximum (FWHM) of $0\farcs8$ (4 spaxels = 0.68~kpc) for the point spread function (PSF). The spectra cover $4750-9300~\rm\AA$ (rest wavelength $\approx4560-8920~\rm\AA$) with a spectral sampling of 1.25 \AA{} and a resolution of R=1750/3750 at the blue/red end of the spectrum.

\subsection{Telluric Corrections}\label{subsec:methods.tellurics}

The oxygen ``B band" telluric feature at $\sim 6880~\rm\AA$ falls atop the \NIIb emission line at the redshift of \gal. To prevent biased measurements of the \NIIb emission line parameters, we apply small corrections to the O$_2$ absorption regions. The atmospheric molecular absorption spectrum is retrieved from the ESO Sky Model Calculator\footnote{\url{https://www.eso.org/observing/etc/bin/gen/form?INS.MODE=swspectr+INS.NAME=SKYCALC}} for Paranal using the parameters for the time of the MUSE observations such as airmass and seasonal value for the precipitable water vapor. We fit the molecular absorption template to the \otwo ``A band" ($\lambda \approx 7590-7700\rm\AA$) to determine the scaling between the template and the observations. For each spaxel in the MUSE datacube with an \otwo A-band signal-to-noise (S/N) ratio of $\geq 3$, we fit the \otwo A-band absorption feature with the template scaled by a constant value and then correct the \otwo B-band region with the scaled template. 

The derived scale factors for each spaxel are near unity ($0.86^{+0.04}_{-0.06}$, median $\pm1\sigma$), as expected for a molecular absorption spectrum tailored for the observing location. We use the known \NII intensity ratio ($\approx 3$, \citealp{storey2000}) as an independent check on our corrections. Before applying the corrections, the \NII intensity ratio was $2.1^{+0.9}_{-0.3}$ for spaxels with $\geq 5\sigma$ detections for both \NIIa and \NIIb. After applying our corrections, the \NII intensity ratio rises to $2.7^{+0.2}_{-0.3}$, within $\sim 1\sigma$ of the expected value, albeit with non-negligible scatter. Any residual issues in the correction are a subdominant source of error as we use the scaled \NIIa flux for our later calculations in \S\ref{sec:nuclei} and \S\ref{sec:galaxy}.

\subsection{Photometric Calibration}\label{subsec:methods.phot}

\begin{table}
    \centering
    \caption{\textit{Gaia} photometry and synthetic photometry from the MUSE datacube for the photometric calibration star to the north-east of \gal. $Gaia$ filter magnitudes ($G$, $Bp$, $Rp$) are converted to conventional filters covered by the MUSE wavelength range ($V$, $r$, $R$, $i$) using the derived photometric relations \citep{riello2018, evans2018}. All photometry has been corrected for line-of-sight reddening $E(B-V)=0.04$ \citep{schlafly2011}. Columns: (1) Filter name, (2) Effective wavelength of the filter, (3) \gaia filter magnitude, (4) MUSE synthetic filter magnitude, and (5) Multiplicative scale factor (see \S\ref{subsec:methods.phot}).}
    \begin{tabular}{lccchc}\hline
         Filter & $\lambda_{\rm{eff}}$ & $Gaia$ & MUSE & $\Delta m$ & Scale  \\
          & [$\rm\AA$] & [mag] & [mag] & [mag] & \\
         \hline
    $G$ & $5836$ & $19.07\pm0.01$ & & \\
    $Bp$ & $5021$ & $19.56\pm0.08$ & & \\
    $Rp$ & $7589$ & $18.03\pm0.03$ & & \\
    $V$ & $5446$ & $19.41\pm0.05$ & $20.07\pm0.01$ & $-0.66\pm0.05$ & 1.84 \\
    $r$ & $6203$ & $19.07\pm0.07$ & $19.54\pm0.01$ & $-0.47\pm0.07$ & 1.54 \\
    $R$ & $6696$ & $18.81 \pm 0.05$ & $19.22\pm0.01$ & $-0.41\pm0.05$ & 1.46 \\
    $i$ & $7673$ & $18.63\pm0.10$ & $18.89\pm0.01$ & $-0.26\pm0.10$ & 1.27 \\
    \hline
    \end{tabular}
    \label{tab:photcal}
\end{table}

A major benefit of IFU observations is reliable flux calibration without needing to account for slit losses. We use the bright star to the north-east of the galaxy to estimate the absolute flux scale. The star is in the Gaia DR2 catalog (ID 4799083000694906368, \citealp{gaia1, gaia2}) and we provide the photometry in Table \ref{tab:photcal}. The \gaia filters extend beyond the MUSE wavelength range, so we convert the \gaia photometry to conventional filters using known photometric relations \citep{riello2018, evans2018}. We extracted the stellar spectrum and computed synthetic magnitudes for Johnson $V$, Johnson $R$, SDSS $r$, and SDSS $i$ using filter transmission curves retrieved from the SVO Filter Profile Service \citep{rodrigo2012}.

Comparing the \gaia and MUSE magnitudes, we notice a wavelength-dependent offset between the predicted and observed magnitudes. We use the magnitude differences to compute a multiplicative scale factor for each filter and use a linear fit to the effective wavelengths and scale factors to place the MUSE datacube on an absolute flux scale. We conservatively estimate a $\sim 10\%$ uncertainty in the absolute flux scale based on the uncertainties in the calculated magnitude differences. 

\subsection{Line Fitting Procedure}\label{subsec:methods.linefitting}

We initially attempted to fit stellar population synthesis (SPS) models to the observed spectra (e.g., STARLIGHT; \citealp{fernandes2003}). However, strong emission lines contaminate the \Ha and \Hb regions and the MUSE wavelength range does not cover the higher-order Balmer lines (e.g., H$\gamma$, H$\delta$) which are usually a better tracer of recent star-formation \citep[e.g., ][]{dressler1999}. Additionally, the continuum emission is too weak to place useful constraints on the stellar absorption features. For these reasons, we choose not to fit SPS models to the observed spectra with the caveat that this introduces a low-level systematic uncertainty to all line flux measurements. We include an additional 10\% uncertainty in quadrature when measuring emission-line fluxes. Instead of SPS models, we use low-order polynomials to estimate the spectral continuum, minimizing the normalized Median Absolute Deviation (nMAD) to prevent strong emission lines from affecting the derived continuum level.

\clearpage
\makeatletter\onecolumngrid@push\makeatother

\begin{figure}
    \centering
    \includegraphics[width=0.9\linewidth]{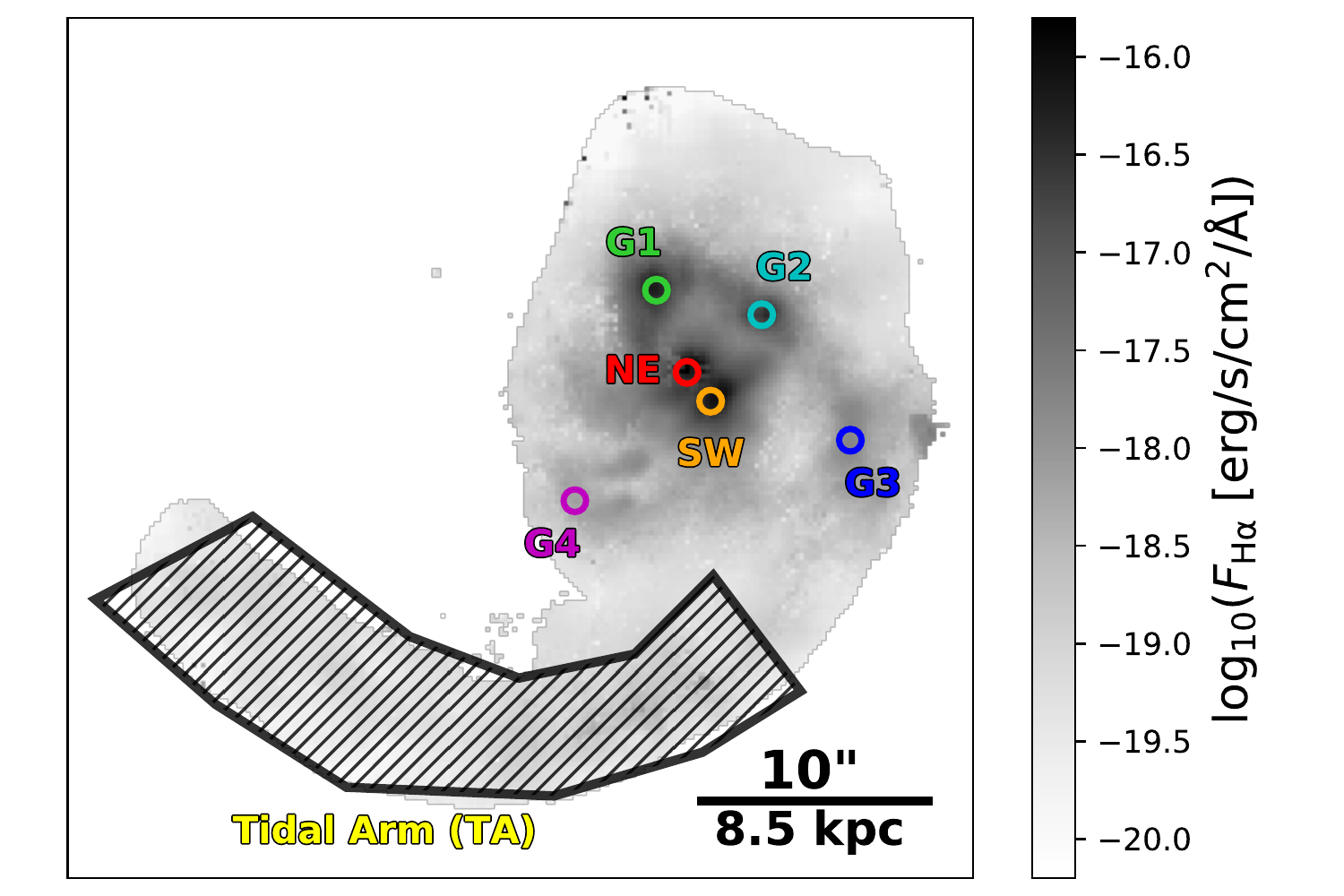}
    \caption{Spatial locations of the extracted spectra used in \S\ref{sec:nuclei} and \S\ref{sec:galaxy} superimposed on a map of the \Ha line flux.}
    \label{fig:speclocs}
\end{figure}

\begin{figure}
    \centering
    \includegraphics[width=0.9\linewidth]{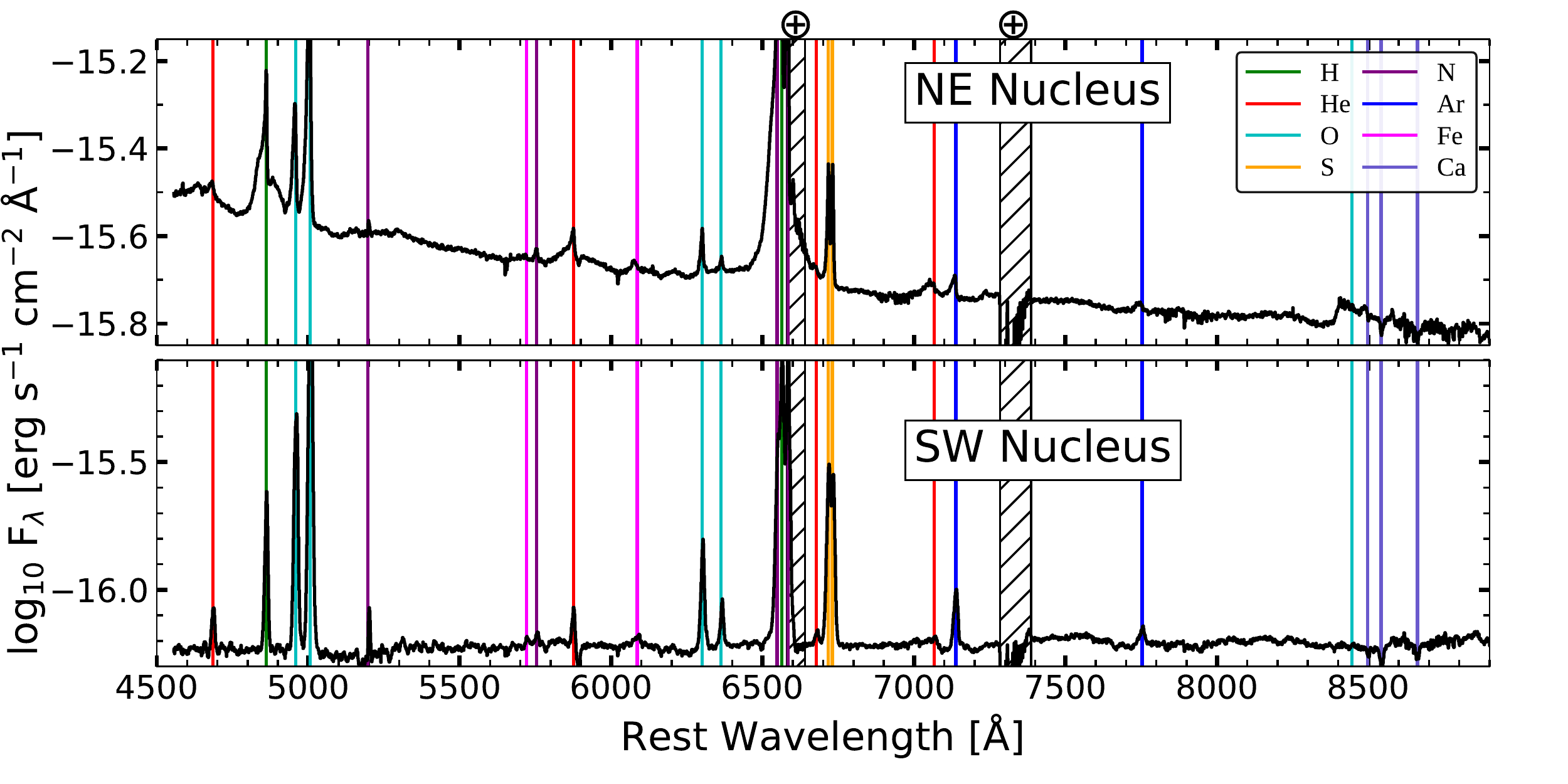}
    \caption{Spectra of the NE (top) and SW (bottom) nuclei. Major emission and absorption features are marked by colored vertical lines. Hatched regions mark the \otwo telluric bands.
    }
    \label{fig:spec-example}
\end{figure}
\clearpage

\makeatletter\onecolumngrid@pop\makeatother

To measure the emission line properties we fit emission-line templates to the continuum-subtracted 1D spectra. Each template has a single velocity shift \vr (relative to the systemic velocity), a velocity width \vfwhm, and includes Gaussian profiles for the following emission lines typical for AGN and active galaxies: \Ha, \Hb, \OIIIlong, \OIlong, \NIIlong, and \SIIlong. The flux amplitudes and velocity parameters are fit simultaneously for each spectrum and uncertainties are estimated from the covariance matrix.

The nuclear region is more complicated, as there are broad emission-line components and asymmetric line profiles. Therefore, we use multiple emission-line templates when fitting the nuclear spectra. To reduce the number of degrees of freedom in the multi-template fitting process, we fix the \OIIIlong flux ratio to 2.99 \citep{storey2000, dimitrijevic2007} for all templates. The \NIIlong lines also have a known flux ratio of $\approx 3$ \citep{storey2000}, but since this region is contaminated by \otwo absorption no constraint is applied. For spectra with broad-line components, we fit the isolated emission lines first (\OIII, \OI, and \SII) to measure the velocity shift and width for each template. The velocity parameters are then held fixed while fitting the \Hb and \Ha + \NII fluxes.

\section{The Nuclei}\label{sec:nuclei}

Fig. \ref{fig:color_images} shows the broad-band (left) and emission-line (right) images for \gal with tidal features to the south-east and north-west of the galaxy center. Fig. \ref{fig:nuc_contours} provides a close-up view of the nuclear region constructed using spectral regions free from strong emission lines. We find two nuclei separated by $\approx 1\farcs7 = 1.4\pm0.1~\rm{kpc}$, consistent with the IR imaging \citep{videla2013, asmus2014}, which we refer to as the `NE' and `SW' nuclei throughout the remainder of this paper. The NE nucleus is unresolved and roughly circular with an axial ratio of $\approx 1$. However, the fainter SW nucleus is resolved along the major axis with a FWHM of $1\farcs1\pm0\farcs1$ but unresolved along the minor axis implying an axial ratio of $\gtrsim 1.4$.

The mid-IR (MIR) locations of both nuclei \citep{asmus2014} are also included in Fig. \ref{fig:nuc_contours}. There is a small ($\approx 1\farcs2$) offset between the MUSE and MIR coordinates for the brighter NE nucleus. This is likely due to uncertainties in the absolute astrometric solution between the two images as neither the MUSE nor the MIR image are calibrated to sub-arcsecond precision. Therefore, we shift the MIR coordinates so that the NE nucleus is aligned to the MUSE image in Fig. \ref{fig:nuc_contours}. This results in the MIR coordinates for the SW nucleus agreeing within $0\farcs3$. The remaining discrepancy is attributed to the extended nature of the SW nucleus at optical wavelengths.

\subsection{Spectra Overview}\label{subsec:nuclei.overview}

\begin{figure}
    \centering
    \includegraphics[width=\linewidth]{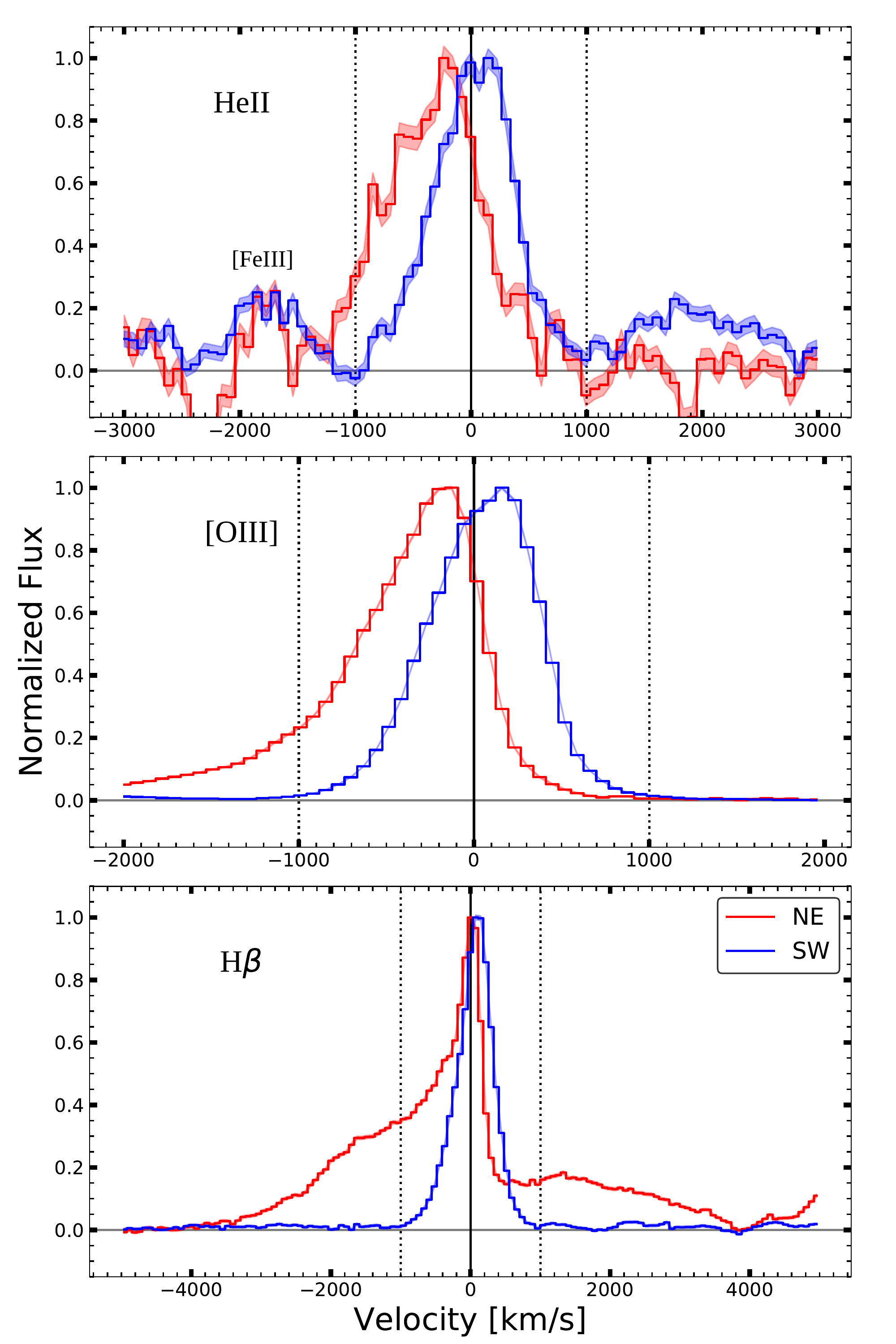}
    \caption{A velocity-space comparison of the emission line profiles for the NE (red) and SW (blue) nuclei, including \HeIIlong (top), \OIIIb (middle), and \Hb (bottom). The vertical dotted black lines in each panel span $\pm1000~\kms$ and serve as a common reference between panels.}
    \label{fig:velcompare}
\end{figure}

\begin{figure*}
    \centering
    \includegraphics[width=\linewidth]{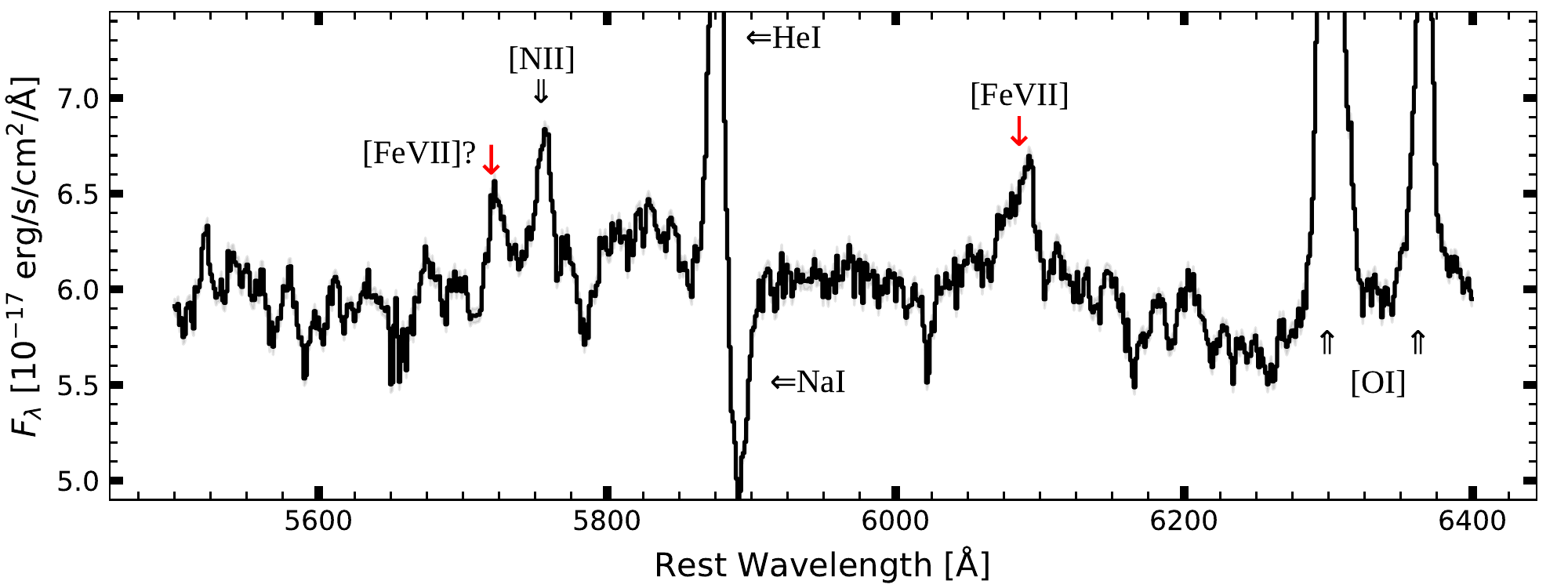}
    \caption{Portion of the spectrum for the SW nucleus from Fig \ref{fig:spec-example} covering the two prominent \FeVII lines. Red arrows indicate the rest wavelength locations for \FeVIIa and \FeVIIb emission lines and black arrows mark other spectral features typical of galaxies and AGN. The \FeVIIb line is detected at high significance, but the \FeVIIa line is stronger and narrower than expected (based on the \FeVIIb line) so we consider the detection of \FeVIIa tentative. 
    }
    \label{fig:FeVIIlines}
\end{figure*}

Fig. \ref{fig:speclocs} shows the spectrum extraction locations for the nuclei relative to the host galaxy along with the non-nuclear regions analyzed in \S\ref{sec:galaxy} with all spectra extracted with an aperture diameter equal to the spatial FWHM ($0\farcs8$). Fig. \ref{fig:spec-example} shows the spectra for both NE and SW nuclei with prominent absoprtion and emission features labeled. The nuclei share many emission lines including strong permitted emission from H and He and forbidden emission lines from \OIII, \OI, [\ion{N}{1}], \NII, and \SII, as is typical of AGN and active galaxies \citep[e.g., ][]{ho1995}. Notable emission lines include \ArIII, which is a reliable metallicity tracer \citep[e.g., ][]{arellano2020, kojima2020}, and \HeIIlong, which can be used as an AGN diagnostic line \citep{shirazi2012}. Notable absorption lines include the \ion{Na}{1} doublet and all three components of the \CaII NIR triplet.

While the nuclei share many of the same spectral features, the velocity parameters of the emission lines differ significantly. Fig. \ref{fig:velcompare} compares the line profiles for \Hb, \OIIIb, and \HeIIlong between the two nuclei. The NE nucleus exhibits strong broad-line emission for \Ha, \Hb, and \ion{He}{1}$\lambda5876\rm\AA$. There are two broad-line emission components, each with $\vfwhm \sim 2\,500 ~\kms$ and separated by $\sim 3\,000~\kms$. The SW nucleus has no evident broad-line emission component but the velocity widths for the emission lines are of order $\approx 700~\kms$, too broad for emission from \HII regions.

Of particular interest is the detection of the coronal \FeVII line in the SW nucleus, as shown in Fig. \ref{fig:FeVIIlines}. The presence of coronal lines imply an abundance of high-energy photons as the $\gtrsim 100~\rm{eV}$ ionization potentials \citep[e.g., ][]{penston1984} require a hard radiation spectrum typically associated with AGN \citep[e.g., ][]{prieto2000, goulding2009}. \FeVII is also observed in the NE nucleus, but the presence of broad emission lines already provides robust evidence for an AGN in the brighter nucleus. 

\begin{figure}
    \centering
    \includegraphics[width=\linewidth]{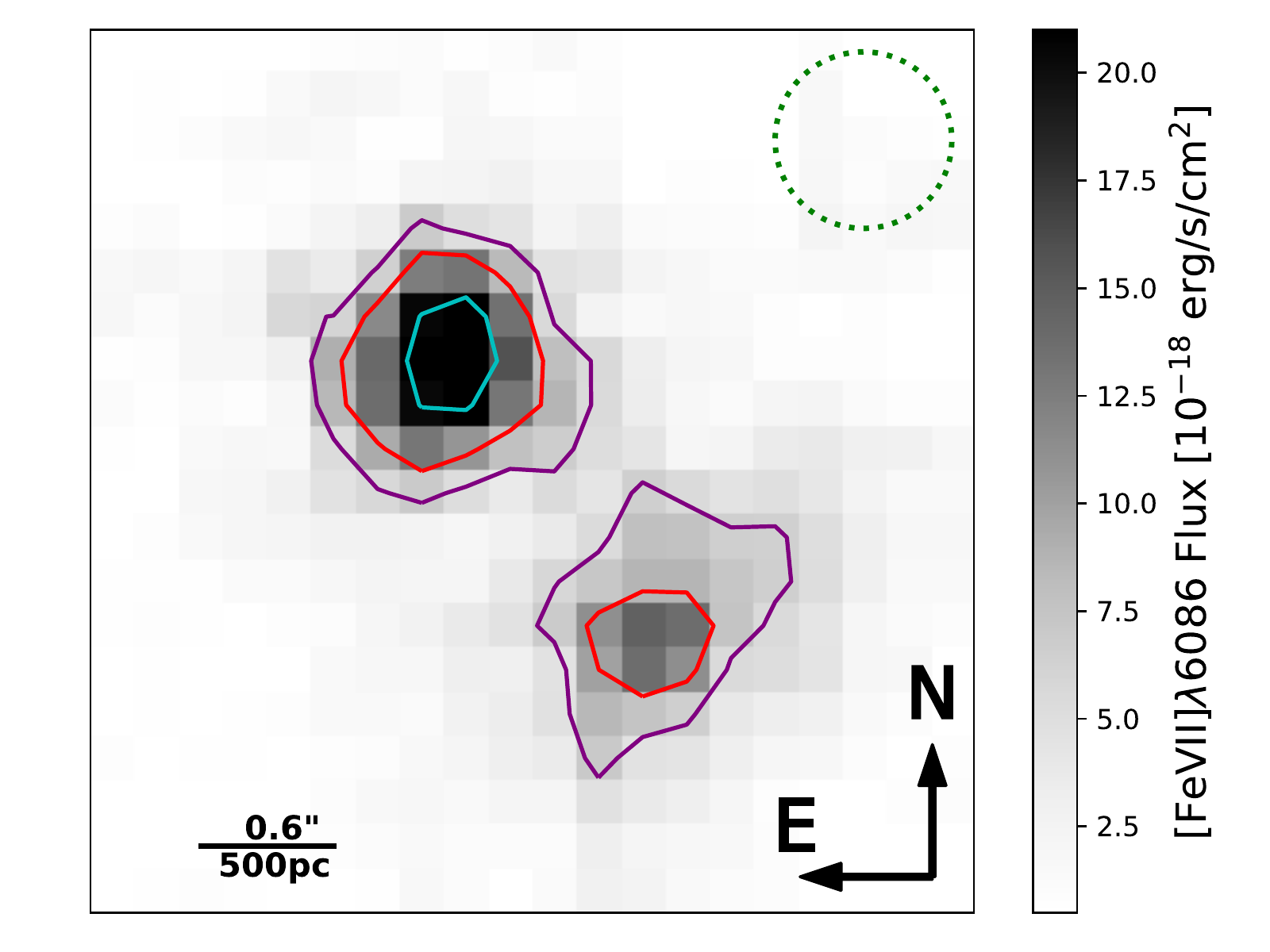}
    \caption{Spatial map of the continuum-subtracted \FeVIIb line flux. Blue, red, and purple contours represent flux levels of $(6, 10, 20)\times 10^{-18}~\rm{erg}~\rm{s}^{-1}~\rm{cm}^{-2}$ and the dotted green circle signifies the spatial FWHM.}
    \label{fig:mapFeVII}
\end{figure}

The detection of \FeVII provides tentative evidence for the presence of an AGN in the SW nucleus. However, the AGN in the NE nucleus also produces high-energy photons which could photoionize gas in the SW nucleus similar to ionization cones seen in other systems \citep[e.g., ][]{wilson1993}. To test this hypothesis, Fig. \ref{fig:mapFeVII} shows the spatial distribution of the continuum-subtracted \FeVIIb line flux. If the \FeVIIb emission in the SW nucleus is produced by photons originating from the NE nucleus, we would expect the \FeVII emission to be extended along the axis connecting the nuclei. Instead, we find similar results to the continuum image in Fig. \ref{fig:nuc_contours}: the NE nucleus is a point-source whereas the SW nucleus is slightly extended to the north-west and consistent with a point-source at $\approx 2\sigma$.

\begin{figure*}
    \centering
    \includegraphics[width=0.95\linewidth]{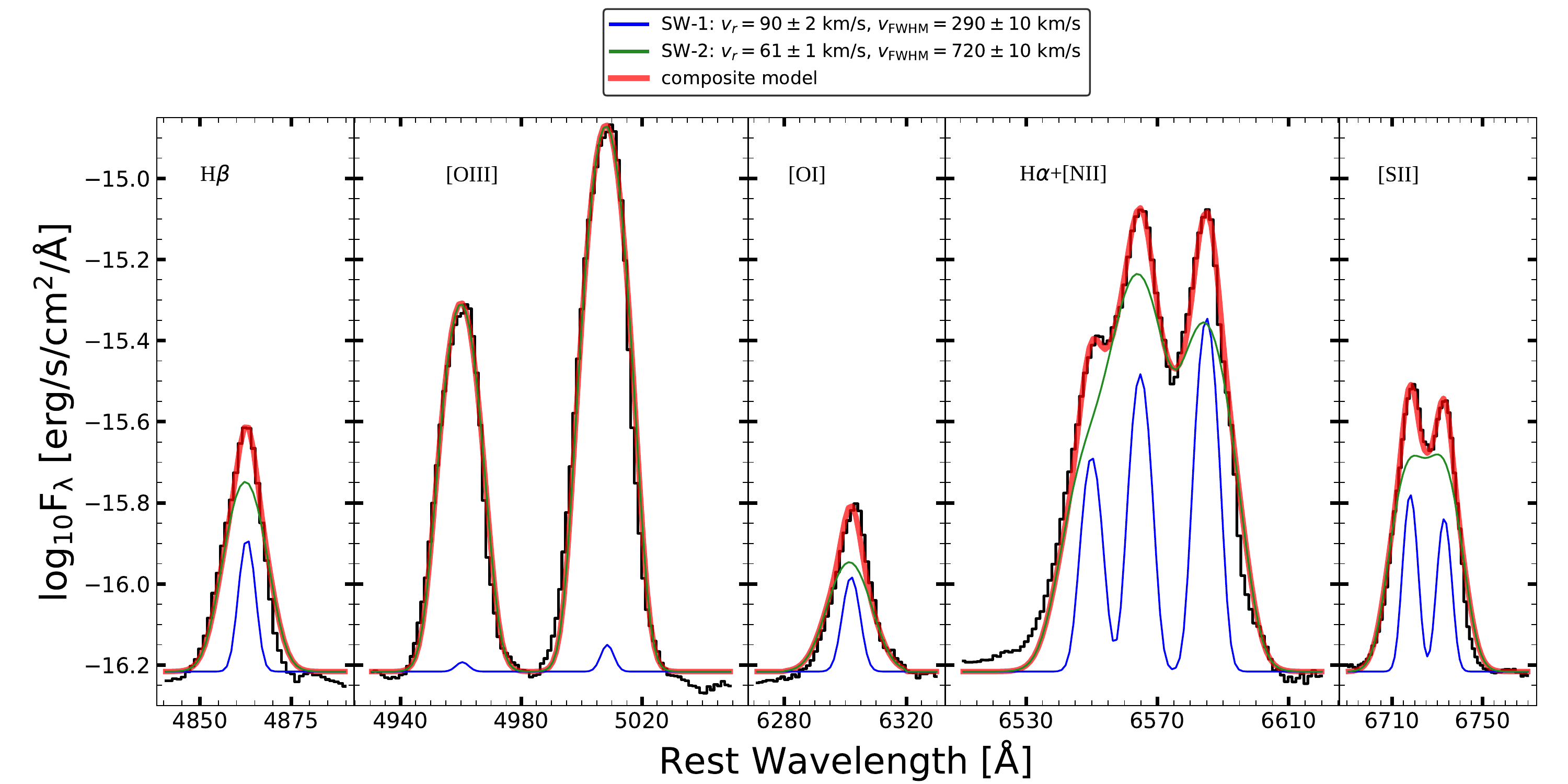}
    \caption{Emission-line decomposition for the SW nucleus. All emission lines are fit simultaneously with the \OIII line ratio fixed (see \S\ref{subsec:methods.linefitting}).}
    \label{fig:SWspecfit}
\end{figure*}

It is possible that the gas densities are too low to produce \FeVIIb between the nuclei in Fig. \ref{fig:mapFeVII}. However, coronal lines are typically detected within $\sim 200~\rm{pc}$ of the AGN \citep[e.g., ][]{prieto2005, mullersanchez2011, amorim2021} whereas the NE and SW nuclei are separated by $\approx 1.4~\rm{kpc}$. Additionally, the observed $\vfwhm \approx 400~\kms$ for \FeVIIb is consistent with coronal-line velocities observed in individual AGN \citep[e.g., ][]{rodriguezardila2011,cerqueiracampos2021}. Therefore, we find compelling evidence that the SW nucleus hosts an AGN. Future adaptive-optics (AO) assisted IFU spectroscopy, especially in the NIR, will place direct constraints on the spatial extent of the coronal-line emission for both nuclei \citep[e.g., ][]{mazzalay2013}.

\subsection{Emission-Line Diagnostics}\label{subsec:nuclei.compBPT}

\begin{figure*}
    \centering
    \includegraphics[width=0.95\linewidth]{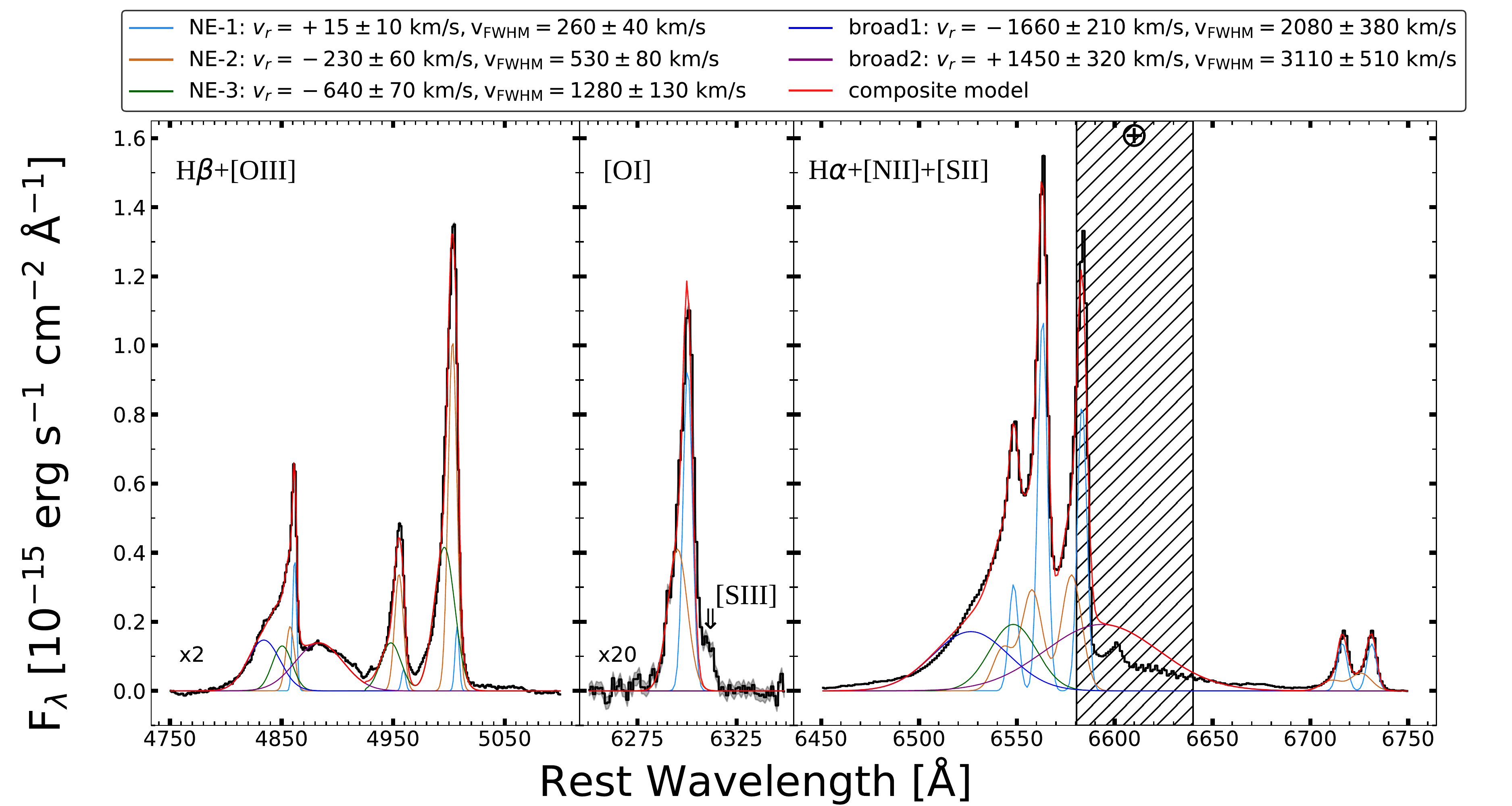}
    \caption{Emission-line decomposition for the NE nucleus. The left and middle panels are scaled by a multiplicative factor given in the lower left for visual clarity. The hatched region marks wavelength regimes contaminated by telluric absorption and are excluded from the fitting process.}
    \label{fig:NEspecfit}
\end{figure*}

As discussed in \S\ref{subsec:methods.linefitting}, we fit the spectra of the NE and SW nuclei with multiple emission-line templates. Fig. \ref{fig:SWspecfit} shows that the SW spectrum is well-fit by two emission templates. A single emission template fails to adequately fit the complex line profiles (e.g., \Hb, \OI) whereas using three templates only leads to marginal improvement. The NE spectrum, shown in Fig. \ref{fig:NEspecfit}, is more complicated. Utilizing only one or two emission templates fails to fit the \OIII profiles and other forbidden emission lines simultaneously, as the asymmetric blue wing of the \OIII profile (see Fig. \ref{fig:velcompare}) ``pulls" the templates away from the line center for \SII, \OI, and \Hb. We find that three emission templates are needed to adequately fit the forbidden emission lines, although the broadest template (NE-3, $\vr \approx -650~\kms$, $\vfwhm \approx 1300~\kms$) contributes almost zero flux for the \OI and \SII emission lines. In addition to the three templates, the NE \Hb line exhibits a complex, double-peaked broad emission profile (Fig. \ref{fig:velcompare}, bottom panel). The blue- and red-shifted broad-line components are each approximated with a Gaussian profile, denoted `broad1' and `broad2' in Fig. \ref{fig:NEspecfit}. 

After fitting each of the emission components for the NE and SW nuclei, we attempt to discern the ionizing source for each nucleus by placing the emission-line ratios on the classical BPT diagrams \citep{baldwin1981,veilleux1987,kewley2001,kauffmann2003}. The BPT classification scheme includes star-formation (SF, sometimes denoted as \HII), AGN, composite (a combination of SF and AGN ionization), and Low Ionization Narrow Emission-line Regions (LINERs). The ionization mechanism for LINERs is still debated, with proposed ionization sources including shocks from merger-induced tidal forces \citep[e.g., ][]{monreal2010, rich2011}, dusty/obscured AGN \citep[e.g., ][]{groves2004, gonzalez2009}, old stellar populations \citep{sarzi2005}, starburst-driven shocks \citep[e.g., ][]{olsson2007}, AGN-driven shocks \citep[e.g., ][]{cheung2016, molina2018}, or some combination of these processes \citep{kewley2006, davies2014, dagostino2019}. We choose not to analyze each template separately as this implicitly assumes the underlying gas velocity distribution. Instead, we sum the flux from all the templates for each line measurement. 

\begin{figure*}
    \centering
    \includegraphics[width=\linewidth]{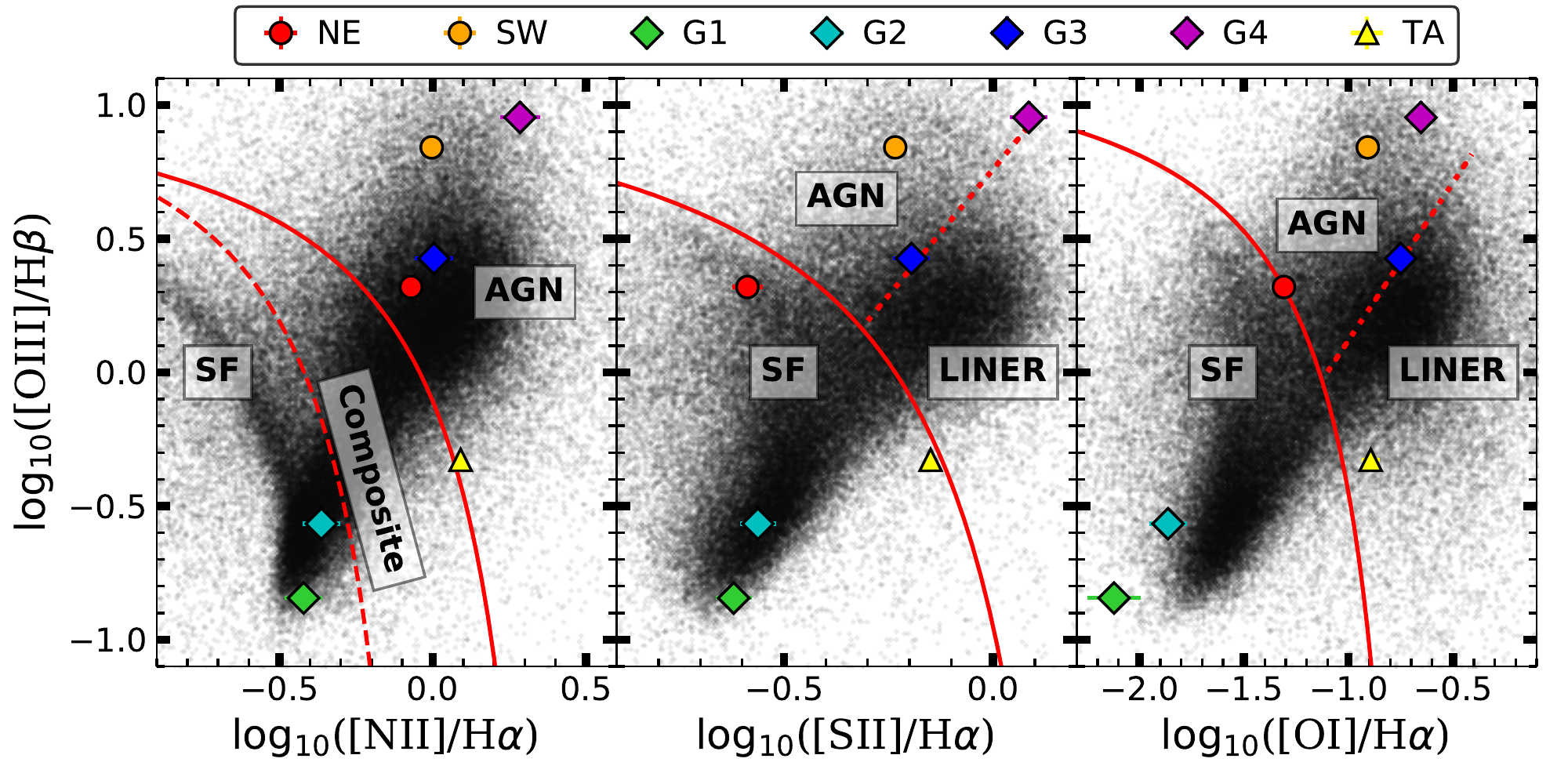}
    \caption{Emission-line diagnostic diagrams for the nuclei (circles) and selected regions within the galaxy (diamonds) compared to SDSS galaxies from the MPA-JHU catalog \citep[black points,][]{york2000, brinchmann2004}. \Hb is undetected in the Tidal Arm spectrum so we place a lower limit on \logOIIIHb signified by the upwards triangle. The red classification lines are taken from \citet{kewley2001} and \citet{kauffmann2003}.}
    \label{fig:fullgalBPT}
\end{figure*}

Fig. \ref{fig:fullgalBPT} shows the line ratios for both nuclei as well as line ratios for the other regions in the galaxy (Fig. \ref{fig:speclocs}) which are discussed in \S\ref{sec:galaxy}. Based on the emission line ratios, the SW nucleus is consistent with AGN photoionization in all three BPT diagrams whereas the NE nucleus falls along the AGN/SF boundary. In addition to the classical BPT diagrams, the nuclei also occupy parameter spaces associated with AGN in the ``$W_{\rm H\alpha}$ vs. \NII/\Ha'' (WHAN) diagram \citep{fernandes2011} and the \HeIIlong classification diagram \citep{shirazi2012}.

Finally, we note that the AGN designation for both nuclei is independent of our choice to analyze the templates collectively. The NE nucleus has broad-line emission components which is robust evidence for an AGN regardless of the emission-line ratios \citep[e.g., ][]{stern2013}. If the SW-1 and SW-2 templates are analyzed separately, the SW-2 template is consistent with AGN photoionization in the emission-line diagnostic diagrams, and exhibits line widths of $\approx 700~\kms$. Conversely, the narrower ($\vfwhm \approx 90~\kms$) SW-1 template has a mixture of SF and LINER classifications. Thus, both nuclei have spectral signatures of AGN regardless of the spectral analysis method. 

\subsection{Modeling the NE Broad-line Emission}\label{subsec:nuclei.models}

The profiles of the broad Balmer lines from the NE nucleus appear double-peaked (Fig. \ref{fig:velcompare}) which is usually interpreted as emission from a relativistic accretion disk around the SMBH \citep[e.g., ][]{chen1989, eracleous2009}. Although the profiles of lines from the surface of an axisymmetric disk are asymmetric because of relativistic effects, a wider variety of profile asymmetries is possible if the disks are non-axisymmetric (e.g., if the streamlines are elliptical or if spiral structure is present in a circular disk). To test if the broad-line components in the NE spectrum can be interpreted as arising from a disk, we fit disk models to the observed broad-line components. Due to the complexity of the narrow-line emission, only the broad-line emission is included in the modeling process and the models do not account for the narrow emission-line components. 

\begin{figure}
    \centering
    \includegraphics[width=\linewidth]{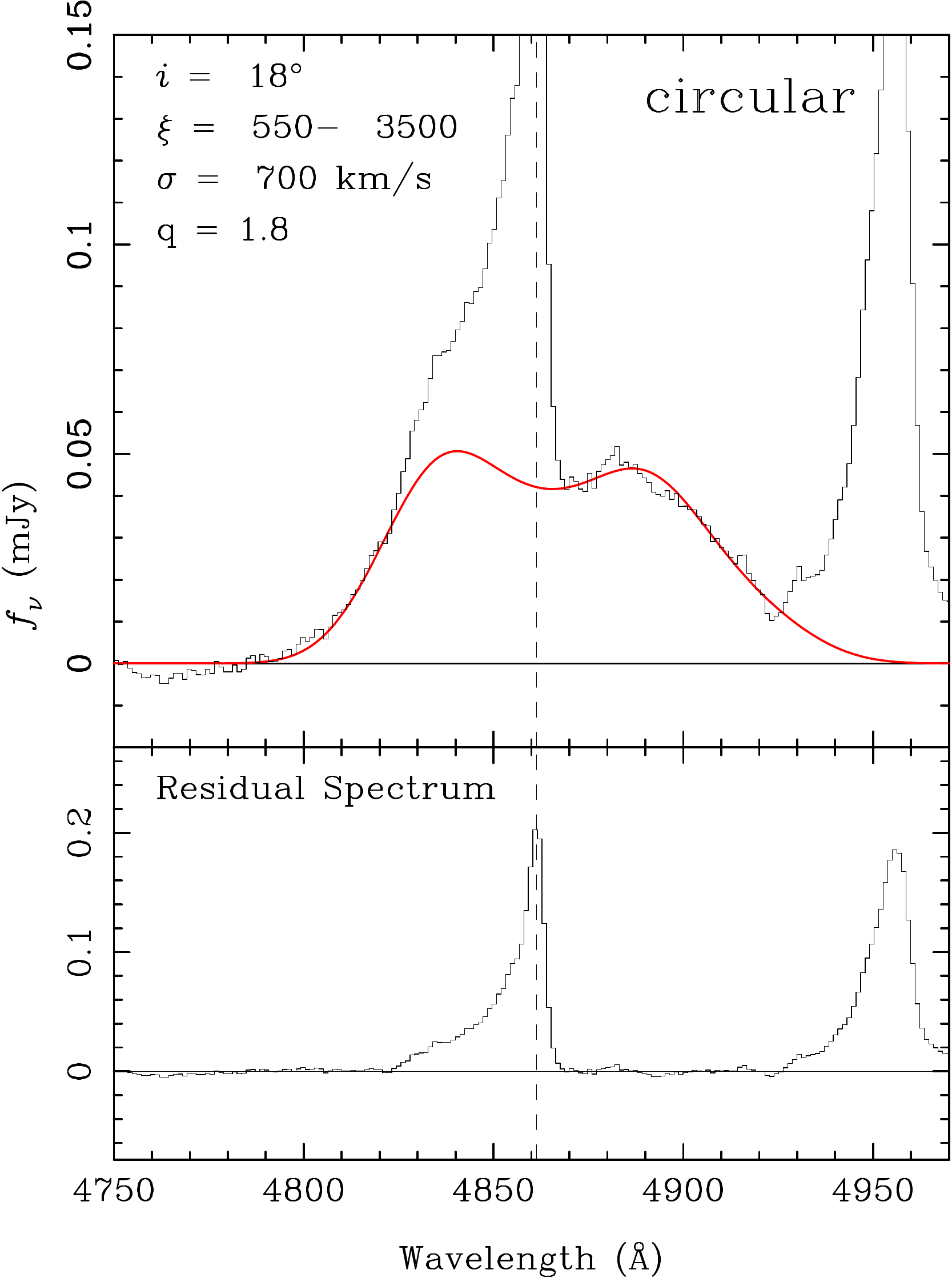}
    \caption{Best-fit broad-line emission profile (top) and corresponding residual spectrum (bottom) for a circular relativistic disk. The narrow emission components are not included in the fitting process due to their complexity and are present in the residual spectrum (bottom panel). The circular disk model does not adequately reproduce the asymmetry of the broad-line emission profile, resulting in a blue-shifted tail in the narrow component of the \Hb line.}
    \label{fig:circular}
\end{figure}

\begin{figure}
    \centering
    \includegraphics[width=0.9\linewidth]{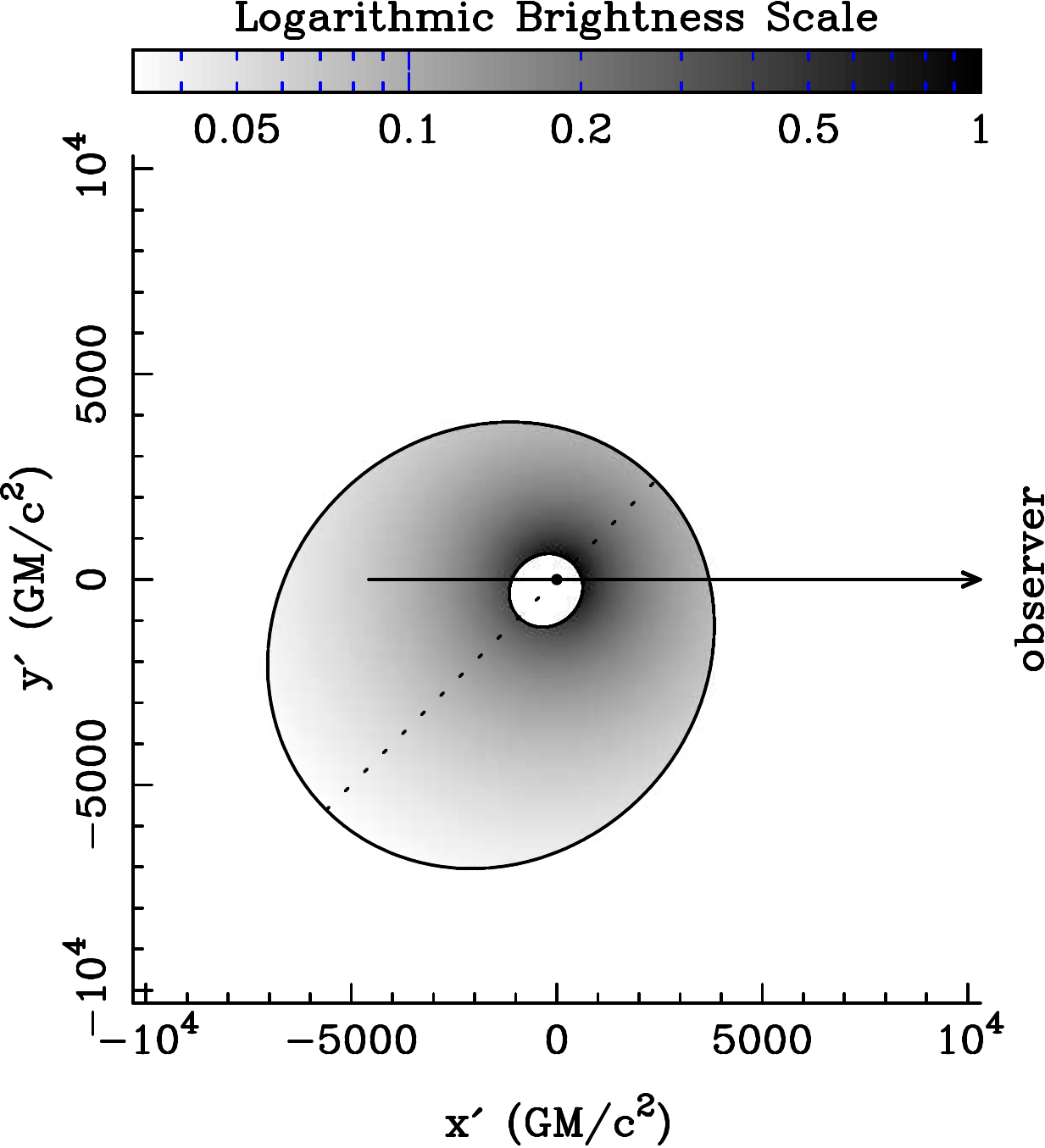}
    \vskip 0.2truein
    \includegraphics[width=0.9\linewidth]{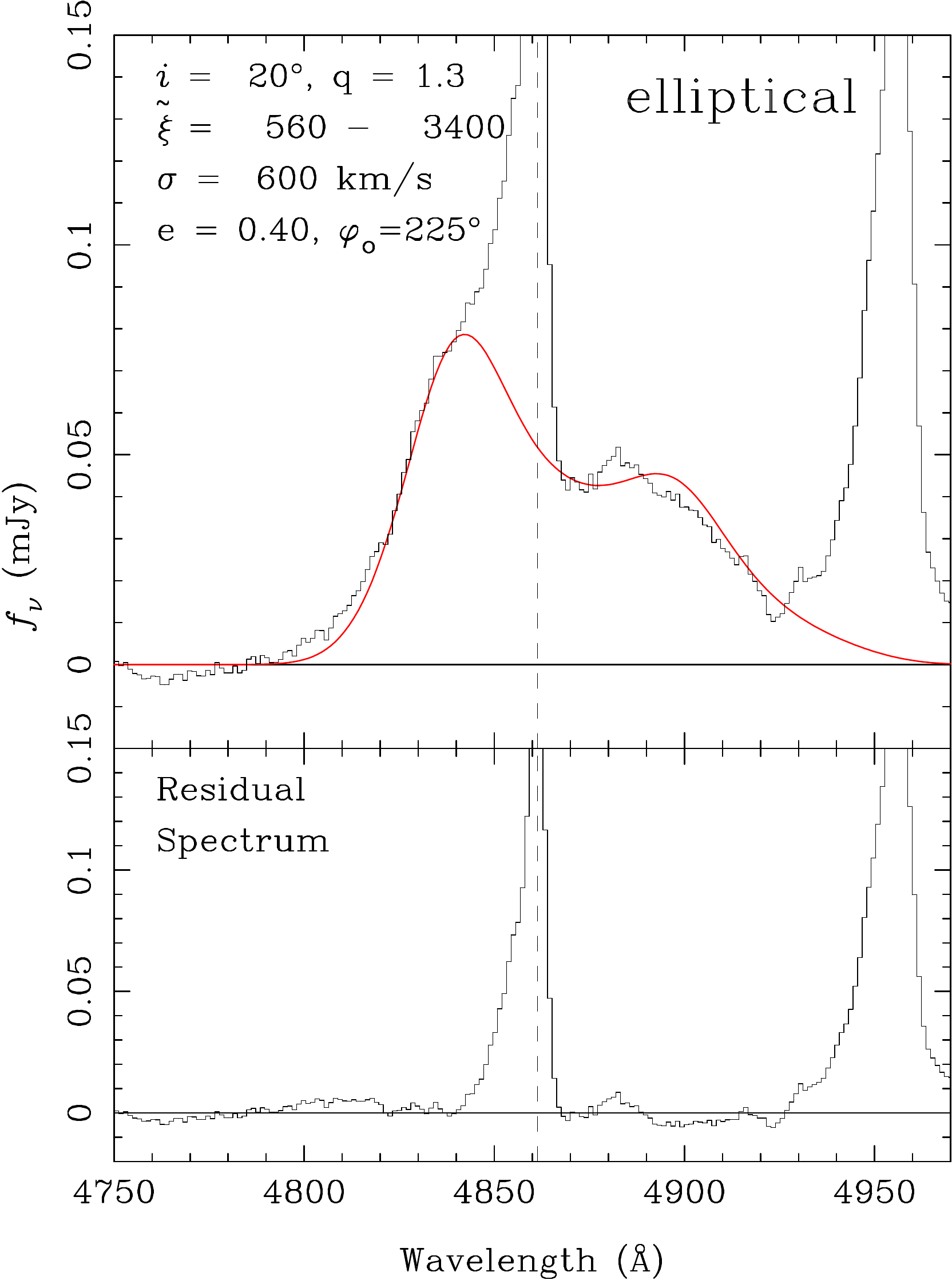}
    \vskip 0.2truein
    \caption{Elliptical-disk model for the double-peaked broad emission lines in the NE nucleus including geometric diagram (top) and the corresponding emission-line profile (bottom). The dashed line in the top panel shows the major axis of the elliptical streamlines. The parameters of the elliptical-disk model are provided in the lower panel and the notation is explained in Section~\ref{subsec:nuclei.models} of the text.}
    \label{fig:ellipdisk}
\end{figure}

\begin{figure}
    \centering
    \includegraphics[width=0.9\linewidth]{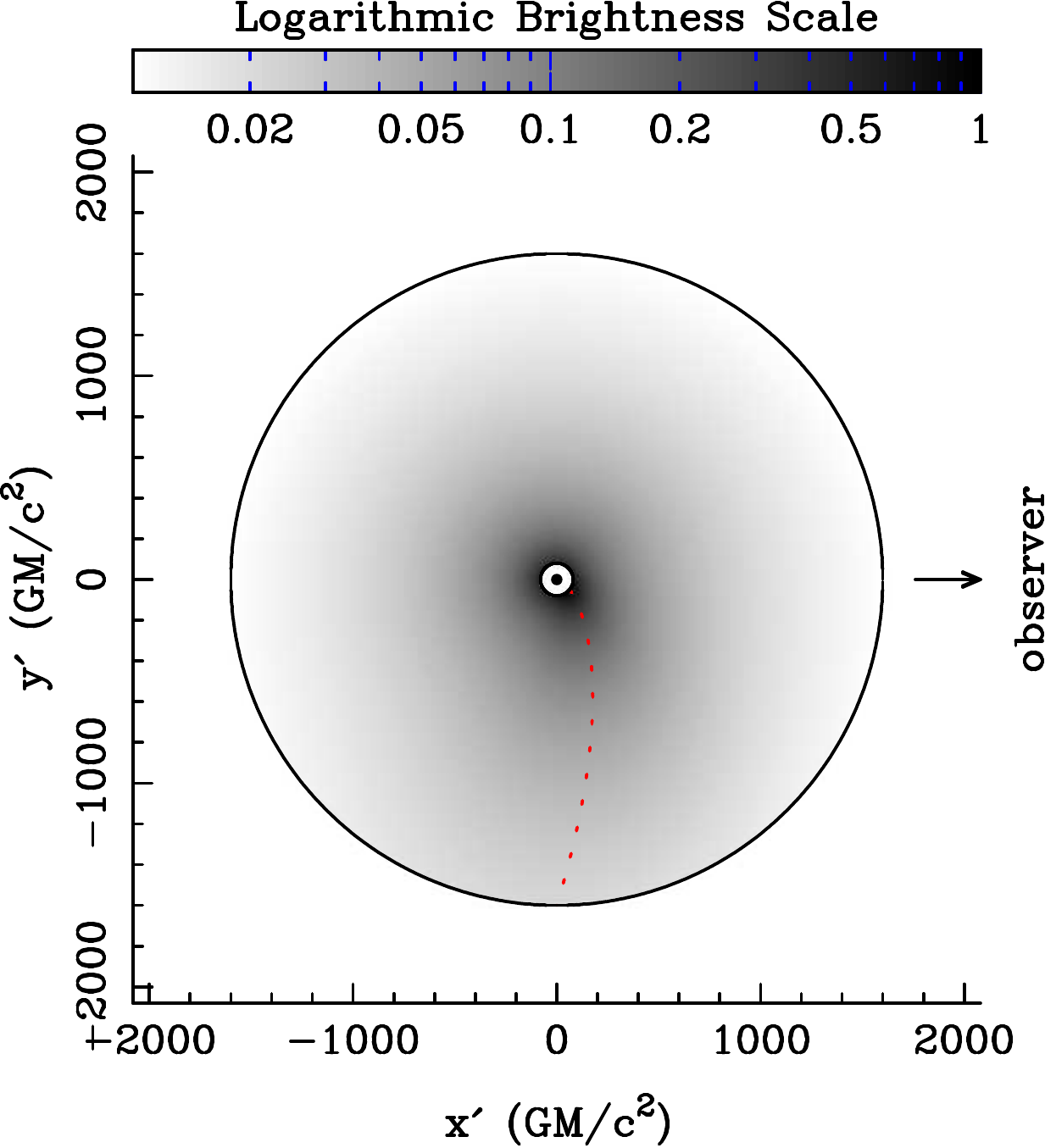}
    \vskip 0.2truein
    \includegraphics[width=0.9\linewidth]{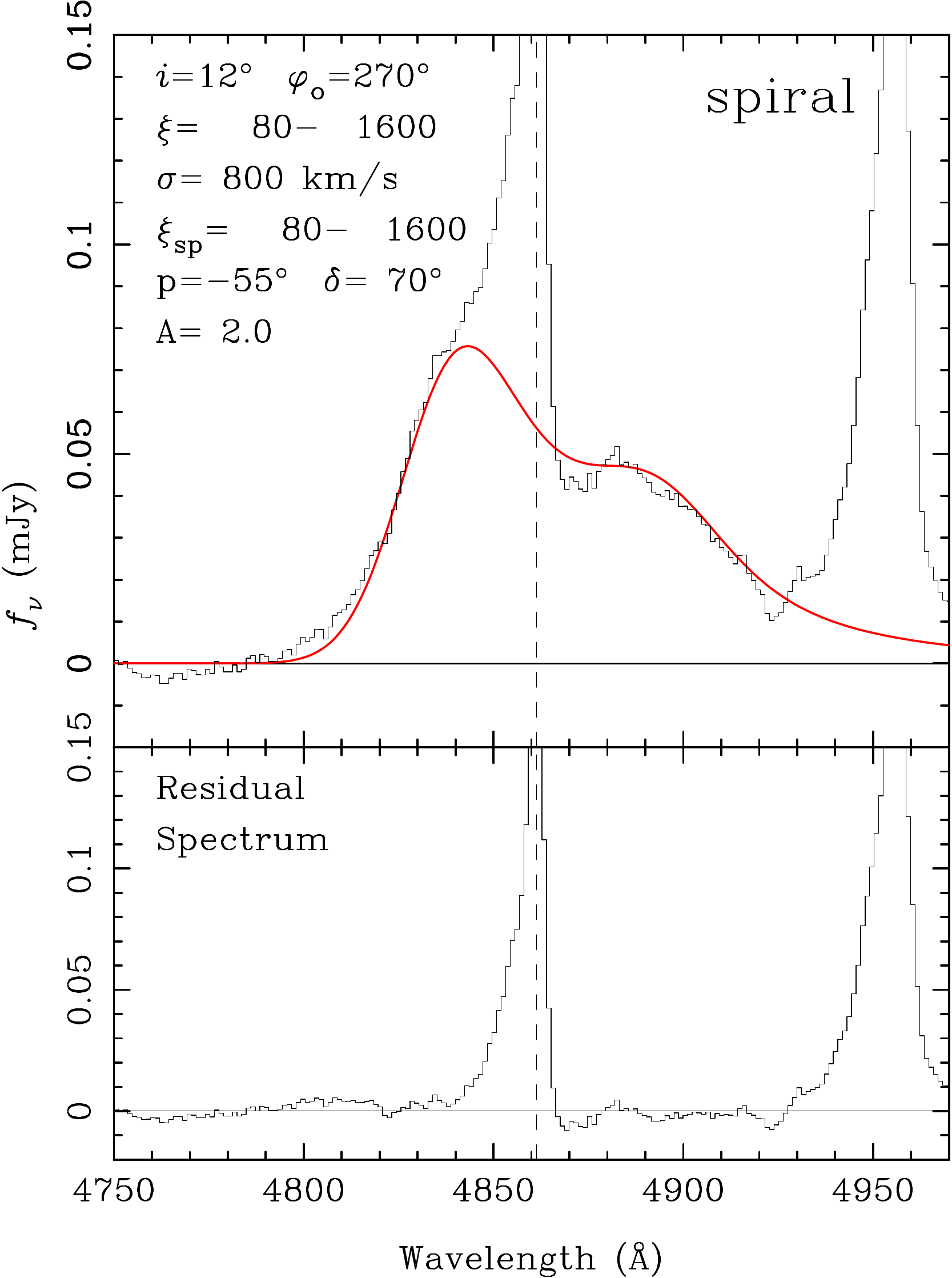}
    \vskip 0.2truein
    \caption{Spiral-arm model for the double-peaked broad-line emission profile in the NE nucleus including geometric diagram (top) and the corresponding emission-line profile (bottom). The spiral arm location is marked by a dotted red line in the top panel. The parameters for the spiral-arm model are provided in the lower panel and the notation is explained in Section~\ref{subsec:nuclei.models} of the text.}
    \label{fig:spiraldisk}
\end{figure}

We first attempted to fit a circular disk model \citep{chen1989} to the broad-line emission components in the NE spectrum. However, Fig. \ref{fig:circular} shows that a circular disk produces a double-peaked broad-line profile with only mild asymmetries. The resulting \Hb residuals have a blue-shifted flux excess out to $\vr \sim -2500~\kms$, far higher than the blue-shifted wings of the \OIII profiles which only extend to $\vr \sim -2000~\kms$. Instead, the asymmetry of the broad-line profile leads to a disk model that is non-axisymmetric with two configurations providing adequate results: an elliptical disk (Fig.~\ref{fig:ellipdisk}) or a circular disk with a spiral arm (Fig.~\ref{fig:spiraldisk}) model.

The elliptical disk model, as described in \citet{eracleous1995}, consists of nested ellipses with their major axes aligned and a fixed eccentricity $e$. The line-emitting portion of the disk extends between inner and outer pericenter distances $\tilde\xi_1$ and $\tilde\xi_2$ (measured in units of the gravitational radius $r_{\rm g}\equiv GM_\bullet/c^2$, where $M_\bullet$ is the mass of the SMBH). The normal to the disk makes an angle $i$ with the line of sight (the inclination angle) and the major axis of the disk makes an angle $\varphi_0$ with the projection of the line of sight in the disk plane (measured counter-clockwise, between apocenter direction and the direction of the observer, as shown in the top panel of  Fig.~\ref{fig:ellipdisk}). The line emissivity of the disk follows a power law with radius of the form $\epsilon\propto r^{-q}$, to describe photoionization of the surface layers by energetic photons from the vicinity of the black hole. The local profile of the emission line is taken to be a Gaussian of velocity dispersion $\sigma$ to describe the effects of electron scattering and/or local turbulent motions. This model has been used to describe emission-line profiles observed in the spectra of TDEs \citep[e.g.,][and references therein]{holoien2019}. It may be appropriate in this case, if a transient disk is being formed out of the debris released by the tidal disruption of a star \citep[i.e., the interpretation favored by][]{payne2020}. In this scenario, the line profile can vary as the structure of the accretion flow changes \citep[see][and references therein]{holoien2019}.

The spiral arm model, described in \citet{gilbert99} and \citet{storchi2003}, has a 1-arm spiral emissivity pattern superimposed on the axisymmetric emissivity pattern of a circular disk. In this model, the axis of the underlying disk makes an angle $i$ with the line of sight, the line emitting portion of the disk lies between radii $\xi_1$ and $\xi_2$ (in units of $r_{\rm g}$), and has an axisymmetric emissivity of the form $\epsilon\propto r^{-q}$. The local line profile is a Gaussian of velocity dispersion $\sigma$. Atop the disk is a 1-arm logarithmic spiral pattern that extends between radii $\xi_{\rm sp,1}$ and $\xi_{\rm sp,2}$ (in units of $r_{\rm g}$) with an angular width $\delta$ and a pitch angle $p$ (the angle between the spiral arm and the outer rim of the disk; $p<0$ for a trailing spiral). The azimuth of the spiral arm, extrapolated to the outer radius of the disk, is $\varphi_0$. The spiral pattern is brighter than the underlying disk by a factor $A$. The spiral arm model has been used to describe the variability of AGN line profiles more generally \citep[e.g.,][]{lewis2010,schimoia17}. Spiral patterns can grow and precess on timescales comparable to a few dynamical times \citep[e.g., ][]{adams89,shu90}, causing the line profile to vary accordingly.

After removing the continuum we fit each of the models to the H$\beta$ profile. The best-fit for each model is determined by visual inspection and we do not attempt an exhaustive search of the parameter space, as this level of detailed modeling is beyond the scope of the present work. However, we show that the disk geometry must be non-axisymmetric if the double-peaked profiles arise from the same emission source.

\section{Galaxy-wide Properties}\label{sec:galaxy}

Considering the unique nature of \asassn and our discovery of a second AGN in the system, we next investigate the surrounding galactic environment. The spectral extraction locations, labeled G1-G4 and Tidal Arm for clarity, are shown in Fig. \ref{fig:speclocs} and coincide with regions of strong line emission in Fig. \ref{fig:color_images}. As described in \S\ref{subsec:methods.linefitting}, each 1D spectrum is fit with a single emission line template after fitting and subtracting a low-order polynomial continuum. The gas kinematics are derived from single-spaxel spectra but we extract specific regions for the BPT analysis to increase the S/N ratio when measuring emission-line parameters. Table \ref{tab:galspec_params} provides the best-fit \vr and \vfwhm values for each non-nuclear location.

\begin{table}[ht!]
    \centering
    \caption{Velocity shift $v_r$ and velocity width $v_{\rm{FWHM}}$ values for each of the non-nuclear regions marked in Fig. \ref{fig:speclocs}.}
    \begin{tabular}{c|cc}\hline
         Location & $v_r$ & $v_{\rm{FWHM}}$ \\
          & [km/s] & [km/s] \\\hline
    G1 & $44 \pm 1$ & $154\pm1$ \\
    G2 & $94\pm 1$ & $136\pm1$ \\
    G3 & $40\pm2$ & $291\pm4$ \\
    G4 & $31\pm 3$ & $247\pm 6$ \\
    Tidal Arm & $42\pm 3$ & $217\pm 6$ \\
    \hline
    \end{tabular}
    \label{tab:galspec_params}
\end{table}

Similar to our analysis of the nuclei, we attempt to discern the ionization source for each location by placing the spectra on the classical BPT diagrams in Fig. \ref{fig:fullgalBPT}. The line ratios for the non-nuclear locations occupy various regions in the diagnostic diagrams suggesting a combination of ionizing mechanisms, as expected for a late-stage merger \citep[e.g., ][]{rich2015}. 

\begin{figure*}
    \centering
    \includegraphics[width=0.95\linewidth]{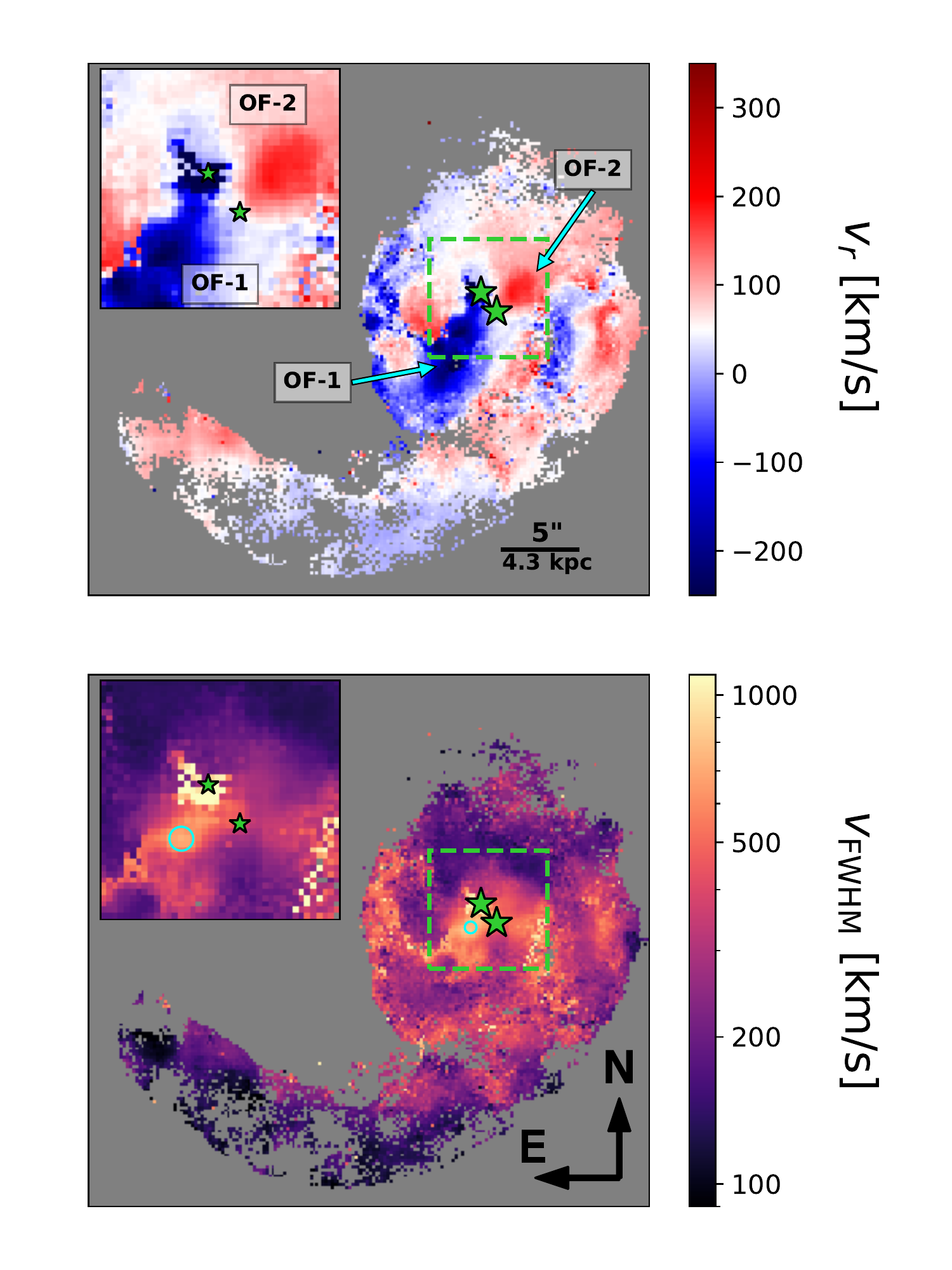}
    \caption{Maps of the mean velocity \vr (top) and line width \vfwhm (bottom). Note the colormap zeropoint for the \vr map has been shifted by $\Delta v = -50~\kms$ to highlight the velocity structure near the nuclei (green stars). The insets highlight the nuclear region inside the dashed green boxes. Two potential outflows, OF-1 and OF-2, are marked in the top panel. The blue circle in the lower panel marks the location of a potential superbubble discussed in \S\ref{sec:galaxy}. \clearpage
    }
    \label{fig:velmap}
\end{figure*}

Fig. \ref{fig:velmap} shows the derived gas kinematics. There is no clear overarching structure in the gas velocity map which would indicate coherent rotation and could be used to infer a dynamical galaxy mass. Instead, there are two potential large-scale outflows near the nuclei (OF-1 and OF-2) and a patchwork of coherent features across the galaxy. OF-1 and OF-2 overlap with the nuclei suggesting an outflow origin but the gas velocities are only a few hundred \kms so gravitational motions cannot be excluded. Future high-resolution spectroscopy covering the \ion{Na}{1} doublet would provide another avenue for studying outflows and bulk gas motions in this system \citep[e.g., ][]{heckman2000, rupke2002, martin2005}.

\begin{figure*}
    \centering
    \includegraphics[width=\linewidth]{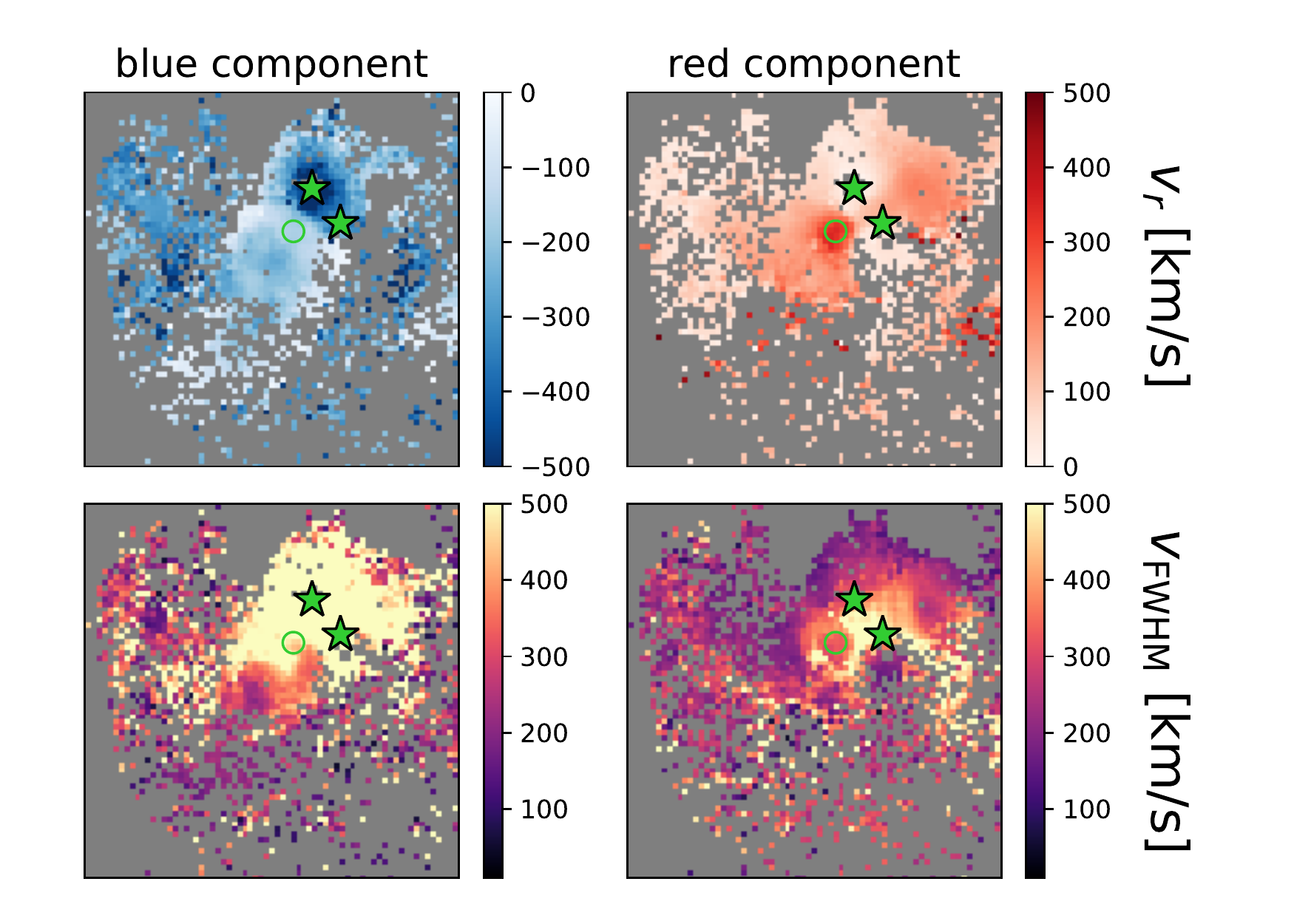}
    \caption{Two-component (left=blue-shifted, right=red-shifted) velocity maps (top: \vr, bottom: \vfwhm) near the nuclei highlighting the possible detection of a superbubble. In each panel, the NE and SW nuclei are marked with green stars and the spectrum extraction radius for the potential superbubble is marked with a green circle. The superbubble appears as a resolved structure in both the \vr and \vfwhm maps for the red-shifted component (right).}
    \label{fig:bubblemap}
\end{figure*}

After fitting the initial velocity maps, we discovered several spaxels with above average \vfwhm in a region south/south-east of both nuclei (Fig. \ref{fig:velmap}, inset). Upon closer inspection, these spaxels exhibited clear multi-component emission features for all major emission lines, including \OIII, \NII, \Ha, and \SII, so we re-fit the spaxels with two emission-line templates. Fig. \ref{fig:bubblemap} shows the results, with the blue-shifted component tracing the large blue-shifted structure seen in Fig. \ref{fig:velmap} (OF-1) whereas the red-shifted component reveals a coherent source to the east of the nuclei. The source is marginally resolved with a spatial FWHM of $1\farcs2\pm0\farcs2$ although the derived spatial FWHM is consistent with a point source at $\approx 2\sigma$.

\begin{figure*}
    \centering
    \includegraphics[width=\linewidth]{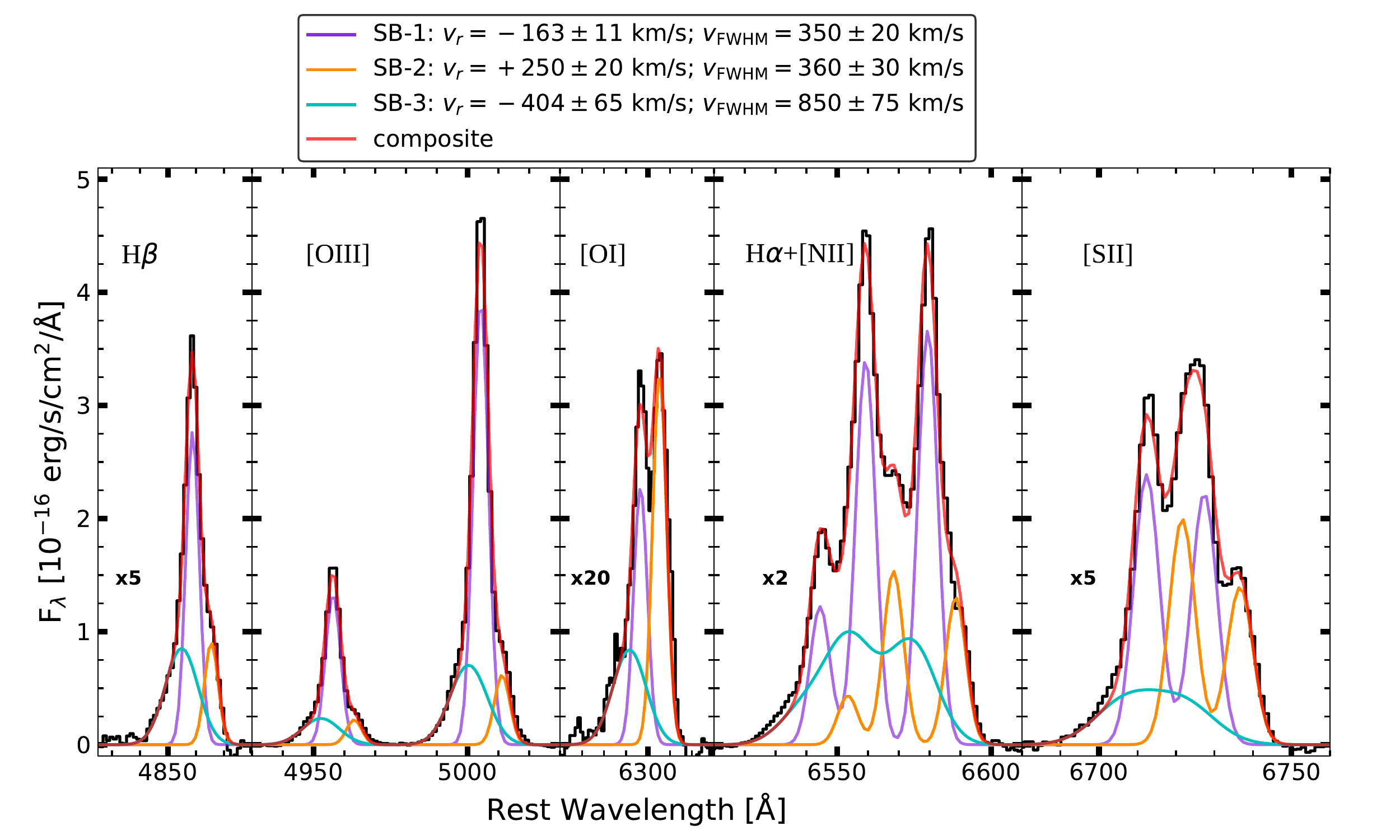}
    \caption{Extracted spectrum and emission decomposition for the potential superbubble structure labeled in Fig. \ref{fig:bubblemap}. Some spectra are scaled by a multiplicative factor for visual clarity with scale factors noted to the left of the corresponding line profile.}
    \label{fig:bubblespec}
\end{figure*}

Fig. \ref{fig:bubblespec} shows the extracted spectrum which is well-fit by three components, two moderately narrow components (SB-1 and SB-2, $\vfwhm \approx 350~\kms$), which are traced by the two-template fitting procedure in Fig. \ref{fig:bubblemap}, and one broader, blue-shifted component (SB-3) with $\vfwhm \approx 750~\kms$. Of these three emission-line templates, SB-2 is the only red-shifted template and has no similar template in either of the NE or SW nuclei. Thus, we consider this a distinct emission source and consistent with a ``superbubble''. Additionally, the SB-2 template exhibits LINER characteristics in the \logOIHa diagnostic diagram consistent with shock-excited emission \citep[e.g., ][]{koo1992, lipari2009}. The \logNIIHa and \logSIIHa classification diagrams are not utilized due to the overlapping emission profiles at these locations and the subsequent uncertainty in the measured line fluxes.  

\section{Implications for ASASSN-14ko}\label{sec:discuss}

Our spatial and spectral analysis of the nuclei and surrounding environment provide new insights into the intriguing nuclear transient \asassn \citep{payne2020} which is located in the NE nucleus \citep{payne2021}. Although we present strong evidence for a second AGN in the system, the SW nucleus has no influence on the evolution of \asassn as the light travel time between the nuclei is greater than the period of \asassn and the sphere of influence of any SMBH is several orders of magnitude smaller than the projected separation of $\approx 1.4~\rm{kpc}$. In this section, we consider several hypotheses for \asassn in the context of our new observations.

Our models of the asymmetric, double-peaked broad-line emission in the NE nucleus indicate that a circular relativistic disk cannot reproduce the observed emission profile (Fig. \ref{fig:circular}) and an non-axisymmetric disk is required. Two models, an elliptical disk (Fig. \ref{fig:ellipdisk}) and a spiral arm superimposed on a circular disk (Fig. \ref{fig:spiraldisk}), produce qualitatively good results. However, they differ significantly in the spatial extent of the broad-line region (BLR): $R_{\rm{BLR}}\approx 5000~r_g$ for the elliptical disk compared to $R_{\rm{BLR}}\approx 1500~r_g$ for the spiral arm + circular disk model.

Differentiating between these models is non-trivial, but each model will have an associated variability timescale. The elliptical disk will precess due to relativistic effects with a period of \citep{eracleous1995} 

\begin{equation}\label{eq:precess}
    P_{\rm{precess}} = 1040 \frac{1+e}{(1-e)^{3/2}} M_8 \xi_3^{5/2}~\rm{years},
\end{equation}

\noindent where $e$ is the disk eccentricity, $M_8$ is the black hole mass in units of $10^8~M_\odot$, and $\xi_3$ is the disk pericenter distance ($\xi_1$ in \S\ref{subsec:nuclei.models}) in units of $1000~r_g$. Using the parameters of the best-fit elliptical-disk model in Fig. \ref{fig:ellipdisk}, $e=0.4$ and $\xi_3 = 0.56$, and a black hole mass of $M_\bullet =  10^7-10^8~M_\odot$ derived by \citet{payne2020}, the associated precession timescale is $\approx 70-700$~years. 

The spiral arm model will evolve more rapidly, with the variability timescale proportional to the pattern speed of the disk. A lower limit on the variability timescale for the spiral arm is the dynamical timescale \citep{storchi2003, lewis2010}

\begin{equation}\label{eq:dyn}
    \tau_{\rm{dyn}} = 200 M_8 \xi_3^{3/2}~\rm{days},
\end{equation}

\noindent where $M_8$ and $\xi_3$ are the same as in Eq. \ref{eq:precess}. For $M_\bullet = 10^8~M_\odot$ and $\xi_3=0.8$ (corresponding to the middle of the disk in Fig. \ref{fig:spiraldisk}), the dynamical timescale is $\tau_{\rm{dyn}}\approx 140$~days, remarkably close to the period of \asassn ($\approx 114$~days). This scenario matches the partial TDE or bound star interpretations for \asassn as each passage of the star would disturb the disk, producing a spiral arm that propagates through the disk and creates the observed asymmetries in the broad emission-line profiles. Even using the outermost edge of the disk, the dynamical timescale is of order $\sim 1$~year and still $\sim 2$ orders of magnitude smaller than the elliptical disk precession period, providing an avenue for distinguishing between these models with future spectroscopic observations.

We searched for disparities between our MUSE spectrum and the spectra of \asassn presented by \citet{payne2020}. The spectra show differences in the shapes and strengths of the emission line profiles. However, we could not reliably remove contamination from the SW nucleus due to ambiguities in the observing setups for the archival spectra (e.g., slit width, slit position angle, and atmospheric seeing). Therefore, we reserve a spectral comparison for future work and reiterate the usefulness of IFU observations for this system.

There is the possibility that the observed broad emission line profiles do not originate from a relativistic disk, in which case the variability timescales discussed above do not apply. Instead, the broad emission-line profiles could stem from separate emission sources such as a binary SMBH system \citep[e.g., ][]{gaskell1983, boroson2009, zheng2016}, where each SMBH hosts an accretion disk and produces one of the broad line profiles, or a biconical outflow \citep[e.g., ][]{zheng1990, zheng1991} where the two lobes of the outflow produce the blue- and red-shifted broad-line profiles. We find these scenarios unlikely, as both theories have repeatedly failed observational tests in other systems \citep[see, e.g., ][]{dietrich1998, fausnaugh2017, runnoe2017, doan2020}, but our current data cannot eliminate them from consideration.

\makeatletter\onecolumngrid@push\makeatother
\clearpage
\makeatletter\onecolumngrid@pop\makeatother

\section{Conclusion}\label{sec:conclusion}

We analyze MUSE data of \gal which exhibits many signatures of a late-stage merger, including a prominent tidal arm and two nuclei. This galaxy also hosts the periodic nuclear transient \asassn \citep{payne2020}, motivating a better understanding of its environment. By combining galaxy-wide properties with an in-depth analysis of the nuclei emission characteristics, we provide several important insights into the environment and characteristics of the merger. 

The spectra of the two nuclei have notable differences in their velocity profiles (Fig. \ref{fig:velcompare}). The NE nucleus, which hosts the periodic nuclear transient \asassn \citep{payne2021}, exhibits asymmetric, double-peaked broad emission lines for \Hb which we model with both axisymmetric and non-axisymmetric relativistic disk models. The broad-line profile cannot be reproduced with a circular disk (Fig. \ref{fig:circular}) but two non-axisymmetric disk models provide good fits to the broad-line profile: an elliptical disk (Fig. \ref{fig:ellipdisk}) or a circular disk with a spiral arm (Fig. \ref{fig:spiraldisk}). It seems inevitable that the inter-outburst line profiles must evolve, due to orbital precession if nothing else (see \S\ref{sec:discuss}). High-quality, phase-resolved spectroscopy over many cycles is required to advance our understanding beyond these initial observations and speculations. However, due to the complex host-galaxy morphology and the presence of dual AGN, care should be taken when planning future observations. The efficacy of slit spectroscopy will depend heavily on the width and orientation of the slit and we suggest IFU observations when possible to alleviate these difficulties.  

The fainter SW nucleus also clearly hosts an AGN. The emission-line ratios for the SW nucleus are consistent with AGN photoionization in diagnostic diagrams (Fig. \ref{fig:fullgalBPT}) and the $\vfwhm \approx 700~\kms$ emission profiles are too broad for star-formation and too luminous for shocks. There is a potential red-shifted outflow originating from the SW nucleus (OF-2 in Fig. \ref{fig:velmap}), suggestive of an accretion source. Finally, the detection of high-ionization lines such as coronal \FeVIIb (Fig. \ref{fig:FeVIIlines}) and broad \HeIIlong (Fig. \ref{fig:velcompare}) require the UV and X-ray continuum emission of an AGN. 

The surrounding galactic environment also exhibits several features of an AGN merger. Velocity maps (Fig. \ref{fig:velmap}) reveal likely outflows from the nuclear region extending for several kpc into the surrounding ISM and a potential AGN-driven superbubble. The host galaxy exhibits many locations with AGN and LINER emission-line ratios, consistent with an AGN merger producing large-scale outflow signatures and shocks. Future high-resolution optical spectroscopy centered on the \ion{Na}{1} doublet and/or radio observations will provide further constraints on outflows, especially for cold gas not probed by our emission-line analyses.

\textit{Facilities}: VLT-MUSE

\textit{Software}: astropy \citep{astropy-ref}, lmfit \citep{lmfit}, mpdaf \citep{MPDAFref, MPDAFref2}, numpy \citep{numpyref}, matplotlib \citep{matplotlib}

\section*{Data Availability}
The raw and reduced MUSE datacubes are publicly available at the ESO Science Archive Facility\footnote{\url{http://archive.eso.org}}.

\section*{Acknowledgements}

We thank the referee for constructive comments and M. Togami for useful discussions. 

MAT acknowledges support from the DOE CSGF through grant DE-SC0019323. BJS, and CSK are supported by NSF grant AST-1907570. BJS is also supported by NASA grant 80NSSC19K1717 and NSF grants AST-1920392 and AST-1911074.  CSK is supported by NSF grant AST-181440. KAA is supported by the Danish National Research Foundation (DNRF132). Support for J.L.P. is provided in part by FONDECYT through grant 1191038 and by the Ministry of Economy, Development, and Tourism’s Millennium Science Initiative through grant IC120009, awarded to The Millennium Institute of Astrophysics, MAS.  L.G. acknowledges financial support from the Spanish Ministry of Science, Innovation and Universities (MICIU) under the 2019 Ram\'on y Cajal program RYC2019-027683 and from the Spanish MICIU project PID2020-115253GA-I00. Parts of this research were supported by the Australian Research Council Centre of Excellence for All Sky Astrophysics in 3 Dimensions (ASTRO 3D), through project number CE170100013. L.G. was funded by the European Union's Horizon 2020 research and innovation programme under the Marie Sk\l{}odowska-Curie grant agreement No. 839090, and partially supported by the Spanish grant PGC2018-095317-B-C21 within the European Funds for Regional Development (FEDER). Support for TW-SH was provided by NASA through the NASA Hubble Fellowship grant \#HST-HF2-51458.001-A awarded by the Space Telescope Science Institute, which is operated by the Association of Universities for Research in Astronomy, Inc., for NASA, under contract NAS5-26555.

Based on observations collected at the European Organisation for Astronomical Research in the Southern Hemisphere under ESO programme 096.D-0296(A).

\bibliography{sample63}{}
\bibliographystyle{aasjournal}

\end{document}